\documentclass{aa}  
\usepackage{natbib}
\usepackage{color}
\usepackage{graphicx}
\usepackage{txfonts}

\begin{document} 

   \title{Diverse reddening distributions in sight lines to type Ia supernovae}
\titlerunning{Diverse distributions of type Ia supernova reddening}

   \author{Lucas Hallgren\inst{1}
          \and Rados{\l}aw Wojtak\inst{1}
          \and Jens Hjorth\inst{1}
          \and Charles L. Steinhardt\inst{1,2}
          }
   \institute{DARK, Niels Bohr Institute, University of Copenhagen, Jagtvej 155, 2200 Copenhagen, Denmark\\
   \email{lucas.o.hallgren@gmail.com, radek.wojtak@nbi.ku.dk}
   \and Department of Physics and Astronomy, University of Missouri, 701 S. College Ave., Columbia, MO 65203 \\
             }

  \abstract
   {
   Accurate cosmological constraints from type Ia supernovae require adequately accurate corrections for host-galaxy extinction. Modelling these corrections is challenged by the problem of disentangling supernova intrinsic colours from host-galaxy interstellar reddening. The latter is commonly modelled in a probabilistic way assuming an exponential distribution $\exp(-E(B-V)/\tau)$ as a universal prior which is applied across all types of supernova host galaxies.
   }
   {We test the robustness of the exponential model of host-galaxy reddening and its universality against predictions based on simulating dust and type Ia supernova distributions in host galaxies of different morphological types.}
   {Our simulations incorporate up-to-date observational constraints on dust masses across host-galaxy morphological types, scaling relations between the dust and stellar disc parameters and the supernova distribution.}
   {
   We find substantial differences between predicted interstellar reddening in late- and early-type host galaxies, primarily driven by the stellar-to-dust mass ratios. The mean simulated reddening in late-type galaxies matches well those derived from type Ia supernova observations, but it is significantly lower for early-type host galaxies. The obtained reddening distributions exhibit an excess of sight lines with zero reddening with respect to the commonly used exponential model, although the difference is quite mild for late-type galaxies. On the other hand, the distribution may peak at $E(B-V)>0$ when considering a population of young type Ia supernovae originating from lower heights within the dust disc of spiral galaxies.
   }
   {
The reddening distribution strongly depends on the supernova host-galaxy morphological type. Assuming a universal reddening prior distribution for modelling peak magnitude-colour relation, which is currently a common practice, gives rise to a spurious scatter in the derived extinction properties. It may also bias relative distances between supernovae originating from different host-galaxy populations. The discrepancy between the simulated reddening in average early-type host galaxies and the observed occurrence of reddened supernovae in these galaxies suggests that reddening does not originate from interstellar dust expected in these hosts.
}


   \maketitle

\section{Introduction}
Type Ia supernovae have long been used to measure cosmological distances and place constraints on cosmological parameters. They played a key role in the discovery of cosmic acceleration and in laying the observational foundation for the concordance $\Lambda$CDM cosmological model \citep{Rie1998,Per1999}. 
Using type Ia supernovae as a cosmological probe was enabled both by the advent of large survey for time-domain astronomy and by developing a range of corrections needed to use them as effective standard candles. The ongoing progress both in terms of observations and standardisation models aims at achieving sufficient accuracy for constraining the dark energy equation of state. However, many challenges are yet to be overcome. The three largest supernova compilations to date -- Pantheon+ \citep{Brout2022}, Union3 \citep{Rubin2023} and DESY5 \citep{DESSN2024} -- used recently in combination with the BAO measurements from DESI \citep{Adame2025} to constrain the dark energy equation of state, result in surprisingly large relative shifts between best-fit cosmological parameters \citep{Cortes2025}. The apparent sensitivity of cosmological constraints to the choice of supernova compilation reflects the magnitude of systematic errors for which 
the most dominant effects tend to bias cosmological results towards evolving dark energy \citep{Dhawan2025}.

The present and future accuracy of measuring cosmological distances with type Ia supernovae relies critically on understanding a range of assumptions 
employed by the standardisation models. Arguably, the most problematic aspect of it is the so-called colour correction which aims to account for 
the observed relation between supernova peak absolute luminosity and its apparent colour. The complexity 
of this relation results from the fact that it is driven by two independent physical effects -- supernova intrinsic emission and host-galaxy extinction -- 
for which the relevant variables are a priori unknown. These two effects can be disentangled in a probabilistic way by modelling type Ia supernova samples. 
This approach was applied to a wide range of supernova data sets, either as a part of light curve analysis \citep[see e.g.][]{Mandel2022,Ward2023,Thorp2022} or as a post-processing analysis of 
supernova colours derived from light curve fitting \citep[see e.g.][]{Mandel2011,Mandel2017,Brout2021,Wojtak2023}. The obtained results, however, depend strongly on the adopted assumptions, in particular the prior 
probability distributions of supernova intrinsic colours and host-galaxy reddening.

Virtually all models of type Ia supernova colours (at a fixed phase of light curves) assume a Gaussian prior for intrinsic colours and an exponential distribution for host-galaxy reddening. 
The former is not motivated by any generic outcome of type Ia supernova simulations, but is rather the easiest choice for a random variable with measurable mean and scatter. The exponential prior for reddening was introduced in early modelling of type Ia supernova data \citep{Jha2007} 
as an approximation of supernova host-galaxy reddening distributions obtained from simulating extinction in disc-like galaxies \citep{riello2005extinction,Hatano1998}. 
However, the exponential model has never been compared quantitatively to the simulated reddening distributions in relation to the assumptions about the spatial density 
of dust and type Ia supernovae. Furthermore, the model does not take into account the fact that about 30 per cent of type Ia supernovae originate from early-type 
galaxies \citep{Hakobyan2012,Pruzhinskaya2020} with very low or negligible interstellar extinction.

The combination of the Gaussian prior for supernova intrinsic colours and the exponential prior for host-galaxy reddening seems to be sufficient to model 
characteristic fat-tailed distributions of supernova apparent colours, well visible in many independent supernova samples \citep{Ginolin2025,Brout2021}. 
However, this does not necessarily guarantee that the model is complete. Based on combined modelling of supernova apparent colours and peak magnitudes, 
many studies find surprisingly low total-to-selective extinction coefficients with $R_{\rm V}\lesssim 2$, either for entire supernova samples \citep{Kessler2009,Burns2014} 
or for supernova population in massive host galaxies \citep{Brout2021,Popovic2021}. These estimates are discrepant with independent and mutually consistent measurements of the extinction coefficient for the Milky Way \citep{Schlafly2016,Fitzpatrick2007,Legnardi2023}, other late-type galaxies \citep{Salim2018,Rino-Silvestre2025} and 
early-type galaxies \citep{Patil2007,Goudfrooij1994}, all pointing to a rather narrow distribution of $R_{\rm V}$ peaking 
at $R_{\rm V}\approx 3$ and decaying fast towards $R_{\rm V}\approx 2$. The most recent improvements in supernova light curve modelling or additional constraints from infrared observations seem to partially alleviate but not eliminate this discrepancy \citep{Thorp2021,Ward2023,Wojtak2023,Thorp2024}. Although one cannot exclude the effect of some special local conditions as the cause of peculiar extinction properties derived from type Ia supernova observations \citep[see e.g.][]{Goobar2008}, 
another possibility is that low extinction estimates arise as a bias due to inaccurate prior distributions assumed for supernova intrinsic colours 
or host-galaxy reddening. The unresolved problem of low extinction estimates derived from type Ia supernova observations affects current 
cosmological measurements including the local Hubble constant determination \citep{Wojtak2024}.

The exponential prior for host-galaxy reddening can be thought of as the most likely model for a positively defined variable given solely its mean value. 
Its two-parameter generalisation given by the $\gamma$-distribution was recently tested against supernova data. The model allows to constrain deviation of the actual probability distribution from the exponential distribution at small reddening. Based on Bayesian hierarchical modelling of supernova light curve parameters, \citet{Wojtak2023} found a preference for distributions assigning maximum probability to sight lines with nonzero reddening. On the other hand, hierarchical Bayesian modelling of supernova light curves showed that a similar trend 
occurs when highly reddened supernovae are excluded, while fitting supernova data without any cuts in supernova colour favours distributions which are more 
peaked for reddening-free sight lines than the exponential model \citep{Ward2023}. These results put into question whether it is justified to use a 
prior reddening distribution that is universal across supernova host-galaxy properties \citep[see also][]{Holwerda2015a,Holwerda2015b} and the related supernova populations, and whether 
the commonly used exponential model is accurate enough to model extinction corrections in type Ia supernova host galaxies.

The goal of this paper is to compare the reddening distributions simulated across supernova host-galaxy morphological types and to quantitatively test the accuracy of the exponential model. Unlike previous studies employing arbitrary scales of dust column densities 
and thus the interstellar reddening \citep{riello2005extinction,Hatano1998}, we compute the absolute scales directly from the expected dust masses based on measurements derived from Herschell infrared observations \citep{smith2012herschel,Cortese2012}. We also 
incorporate a wide range of updates on the spatial distribution of dust and scaling relations between the dust and stellar disc parameters. They were obtained from studies of dust emission exploiting comprehensive observational data from the DustPedia database \citep{Davies2017}.

The outline of the paper is as follows. In section~\ref{sect:sim} we describe the simulations and the observational foundation for the adopted 
assumptions about the spatial distribution of dust and type Ia supernovae across supernova host-galaxy morphological types. The results of the simulations, 
as well as the quantitative comparison between the simulated distributions and the exponential model are presented in section~\ref{sect:res}. 
In section~\ref{sect:disc} we discuss our results and draw implications for possible improvements of modelling type Ia supernova colours and 
host-galaxy extinction.
   
\section{Simulations}
\label{sect:sim}

\begin{table*}
\caption{Parameters adopted to simulate distributions of dust and type Ia supernovae in a typical late-type (LT) or S0-type galaxy.
}
\label{table:disky_setups}     
\centering          
\begin{tabular}{ l c c}     
\hline\hline       
     Parameter                 & Value                                  & Scaling relation                                                          \\
 \hline
 Total stellar mass [LT,S0]        & $\log_{10}(M_{\star}/M_{\odot}) = [10, 10.5]$                 &         N/A                                                     \\
 Optical radius [LT,S0] & $R_{25}=[8.5, 11.5]\,{\rm kpc}$ & $\log_{10}(R_{25}/{\rm kpc})=0.26\log_{10}(M_{\star}/M_{\odot})-1.67\,^{\rm a}$  \\
\hline
 \multicolumn{3}{c}{SN Ia disc parameters}\\ 
\hline                    
 Radial scale [LT,S0]              & $R_{Ia}\equiv R_{\star}=[2.1, 2.9]\,{\rm kpc}$                  & $R_{\star}=0.25R_{25}\,^{\rm b}$                                 \\
 Vertical scale [LT,S0]            & $h_{\rm Ia} = [0.47, 0.63]\,{\rm kpc}$                 &                        $h_{\rm Ia} = 0.055R_{25}\,^{\rm c}$                          \\
 Radial truncation         & $R_{25}$         &                      N/A                                     \\
 Vertical truncation       & $6h_{\rm Ia}$        &                         N/A                           \\
\hline
 \multicolumn{3}{c}{SN Ia bulge parameters}\\ 
\hline
  Bulge-to-total SN rate [LT,LT,S0]        & $N_{\rm b}/N_{\rm t}\equiv M_{{\rm b},\star}/M_{\star} = [0,\,0.3,\,0.5]$            &  N/A                                      \\
  Effective radius [LT,S0]          & $R_{\rm e,Ia}\equiv R_{{\rm e},\star} = [0.43, 0.57]\,{\rm kpc}$ & $R_{\rm e}/R_{\star}=0.2\,^{\rm d,e}$                 \\
  S\'ersic index & $n=2$ & N/A \\
  Truncation radius         & $8R_{\rm{e,Ia}}$      &                      N/A                            \\
\hline
 \multicolumn{3}{c}{Dust disc parameters}\\ 
\hline
  Total dust mass [LT,S0]           & $\log_{10}(M_{\mathrm{d}}/M_{\odot}) = [6.9, 6.1]$       & $\log_{10}{(M_{\mathrm{d}}/M_{\star})} = [-3.1, -4.4]\,^{\rm f,g}$         \\
  Radial scale [LT,S0]    & $R_{\rm d} = [3.0, 4.0]\ \mathrm{kpc}$             & $R_{\rm d}/R_{\star}=1.4\,^{\rm h}$      \\
  Vertical scale [LT,S0] & $h_{\rm d} = [0.26, 0.35]\ \mathrm{kpc}$             & $h_{\star}/h_{\rm d}=1.8\,^{\rm h}$      \\
  Radial truncation         & $2R_{25}\,^{\rm i}$     &                   N/A                                         \\
  Vertical truncation       & $ 6h_{\rm d}$          &                       N/A                           \\
 \hline\hline                  
\end{tabular}
\tablefoot{The dust distribution is given by an exponential disc and the supernova distribution is a sum of an exponential disc and a spherical bulge with the density profile given by the S\'ersic model. The model includes three cases of the bulge-to-total supernova rate ratios: 0.,0.3, 0.5, where the latter corresponds to the S0-type galaxy. Subscripts $\{{\rm d},\star,{\rm b},{\rm Ia}\}$ denote respectively dust, stars, bulge and type Ia supernovae. $R$ and $h$ are the radial and vertical disc scales, $R_{\rm e}$ is the effective radius of the S\'ersic density profile.\\
References: $^{\rm a}$\cite{Pilyugin2021},$^{\rm b}$\cite{casasola2017radial},$^{\rm c}$\cite{hakobyan2017supernovae},$^{\rm d}$\cite{Laurikainen2010},$^{\rm e}$\cite{Graham2008},$^{\rm f}$\cite{Cortese2012},$^{\rm g}$\cite{smith2012herschel},$^{\rm h}$\cite{Xilouris1999},$^{\rm i}$\cite{smith2016far}}
\end{table*}

We simulate supernova host-galaxy reddening using Monte Carlo method based on sampling supernova 
positions in host galaxies and computing dust column density $\sigma_{\rm dust}$ along random sight 
lines. The reddening $E(B-V)$ is derived from the extinction $A_{\rm V}$ in the $V$-band which is 
computed as
\begin{equation}
    A_{\rm V}=[(A_{\rm V}/N_{\rm H})(M_{\rm H}/M_{\rm dust})/m_{\rm H}]\sigma_{\rm dust},
\end{equation}
where $A_{\rm V}/N_{\rm H}$ is the extinction per unit column density of hydrogen, which 
depends on the microscopic model of dust composition, $M_{\rm H}/M_{\rm dust}$ is 
the gas-to-dust mass ratio and $m_{\rm H}$ is the atomic mass of hydrogen. We assume a Milky Way dust 
composition and its model developed by \citet{Weingartner2001ApJ} for which $A_{\rm V}/N_{\rm H}=5.3\times 10^{-22}\,{\rm mag}\,{\rm cm}^{-2}\,{\rm H}^{-1}$ and $R_{\rm V}=A_{\rm V}/E(B-V)=3.1$ \citep{Draine2007}. 
We adopt a gas-to-dust mass ratio $M_{\rm H}/M_{\rm dust}=10^2$ which is a typical value measured 
across all galaxy morphological types \citep{Casasola2020}.

We consider four broad classes of host-galaxy morphological types which differ in terms 
of the underlying spatial distributions of both supernovae and dust. We assume that type Ia 
supernovae trace the stellar mass. This assumption is supported by a number of studies showing 
that type Ia supernova positions are most closely associated with the host-galaxy light distribution in optical bands \citep{Andersen2015,Pritchet2024}. The scale parameters of the stellar and dust distributions are given by observational scaling relations, and they are described in detail below for each host-galaxy morphological type. Tables~\ref{table:disky_setups}-\ref{table:mprph_setups} provide a concise overview of 
all parameters and the underlying scaling relations, respectively, for late-type (including transitional lenticular S0-type) 
and early-type supernova host galaxies. The galaxy models are computed assuming fixed stellar masses representing typical supernova host galaxies: $10^{10}M_{\odot}$ for late-type galaxies, and $10^{10.5}M_{\odot}$ for S0 and elliptical galaxies. We generate $5\times 10^{4}$ random realisations of type Ia supernova positions per host and $20$ random sight lines per supernova.

As a more exploratory part of our study, we consider the possible impact of environmental differences between prompt (following star formation history with typical delays of $0.1-0.5$~Gyr) and delayed (exploding $\gtrsim 0.5$Gyr after star formation) populations of type Ia supernovae \citep{Mannucci2005,Scannapieco2005,Maoz2012}. In this scenario, one may expect that the two supernova populations have different spatial distributions due to the fact that prompt supernovae are more associated with star-forming regions, and delayed supernovae with old stellar environments. The difference
may be primarily visible in the extent of supernova vertical positions, where a narrower range is expected for the prompt population. Tentative evidence for this effect was found from an analysis of type Ia supernova positions in edge-on galaxies \citep{Barkhudaryan2023}. We explore the impact of supernova vertical positions on the reddening distribution in section~\ref{sect:layers}.

\begin{table}
\caption{Parameters adopted to simulate distributions of dust and type Ia supernovae in a typical elliptical (E) galaxy.
}            
\label{table:mprph_setups}     
\centering          
\begin{tabular}{ l c c  }     
\hline\hline       
     Parameter                       & Value                                          & Relation                                                                   \\
 \hline
  Total stellar mass              & $\log_{10}(M_{\star}/M_{\odot}) = 10.5$                         &                         N/A                                             \\
  Effective radius               & $R_{{\rm e},\star} = 3.25\,\mathrm{kpc}$                  & $R_{{\rm e},\star}(M_{\star})\,^{\rm a}$                                               \\
  S\'ersic index & $n=4$ & N/A        \\
\hline  
 \multicolumn{3}{c}{SN Ia parameters}\\ 
\hline  
  Effective radius                & $R_{\rm e,Ia} = 3.25\,\mathrm{kpc}$      &  $R_{\rm e,Ia}\equiv R_{{\rm e},\star}$                  \\
 Truncation radius               & $8R_{\rm e,Ia}$              &                   N/A                                    \\
\hline
 \multicolumn{3}{c}{Dust parameters}\\ 
\hline
  Total dust mass  & $\log_{10}(M_{\mathrm{d}}/M_{\odot}) = 4.7 $ & $\log_{10}{(M_{\mathrm{d}}/M_{\star})} =$    \\
     &  &  $= -5.8\,^{\rm b}$    \\
  Effective radius   & $R_{\rm e,d} = 3.25\,\mathrm{kpc}$              & $R_{\rm e,d} \equiv R_{{\rm e},\star}$                                                                 \\
 Truncation radius  & $8R_{\rm e,d}$                & N/A                                                           \\
\hline\hline                  
\end{tabular}
\tablefoot{Both the dust and type Ia supernovae trace the stellar distribution with the density profile approximated by the S\'ersic model. The two rightmost columns list the relevant scaling relations used to estimate the parameters based on the assumed stellar mass $M_{\star}=10^{10.5}M_{\odot}$. Subscripts $\{{\rm d},\star,{\rm Ia}\}$ denote dust, stars, and type Ia supernovae. $R_{\rm e}$ is the effective radius of the S\'ersic density profile.\\
References: $^{\rm a}$\cite{hyde2009curvature},$^{\rm b}$\cite{smith2012herschel}}
\end{table}

\subsection{Late-type galaxies}

We model the supernova distribution as the sum of disc and bulge components. We assume an exponential disc model with the density distribution given by
\begin{equation}
    \rho_{\rm disc,Ia}\propto \exp\Big(-\frac{R}{R_{\rm Ia}}-\frac{|z|}{h_{\rm Ia}}\Big),
\end{equation}
where $(R,z)$ are the radius and height in cylindrical coordinates, $R_{\rm Ia}$ and $h_{\rm Ia}$ are the radial and vertical disc scales. We model the bulge component using the S\'ersic model \citep{Sersic1963,Sersic1968} for which the surface 
density is 
\begin{equation}
    \Sigma_{\rm bulge,Ia}\propto\exp\Big[-b_{\rm n}\Big(\frac{R_{\rm p}}{R_{\rm e,Ia}}\Big)^{1/n}\Big],
\end{equation}
where $R_{\rm p}$ is the projected radius, $R_{\rm e,Ia}$ is the effective radius defined as the radius of the aperture containing a half of the projected cumulative distribution, $n$ is the S\'ersic index and $b_{\rm n}$ is a fixed function of $n$ derived from the definition of $R_{\rm e}$. In order to model spatial distributions, we use an analytic approximation to the 3D deprojected density profile from \citet{Prugniel1997}. The relative rates of type Ia supernovae originating from the bulge and disc component are assumed to be equal to the corresponding bulge-to-disc stellar mass ratio. We consider two cases of bulge-to-total mass ratios with $M_{\rm b,\star}/M_{\star}=[0.,0.3]$ (hereafter referred to as $B/T$). The adopted values span a range that includes the majority of $B/T$ ratios measured from observations \citep{Weinzirl2009,Laurikainen2007,Pastrav2021}. We assume a pseudo-bulge with $n=2$ as a typical S\'ersic index measured in galaxies of late or transitional morphological types \citep{Laurikainen2007}.

\begin{figure*}[htbp]
    \centering
    \includegraphics[width=1.1\linewidth]{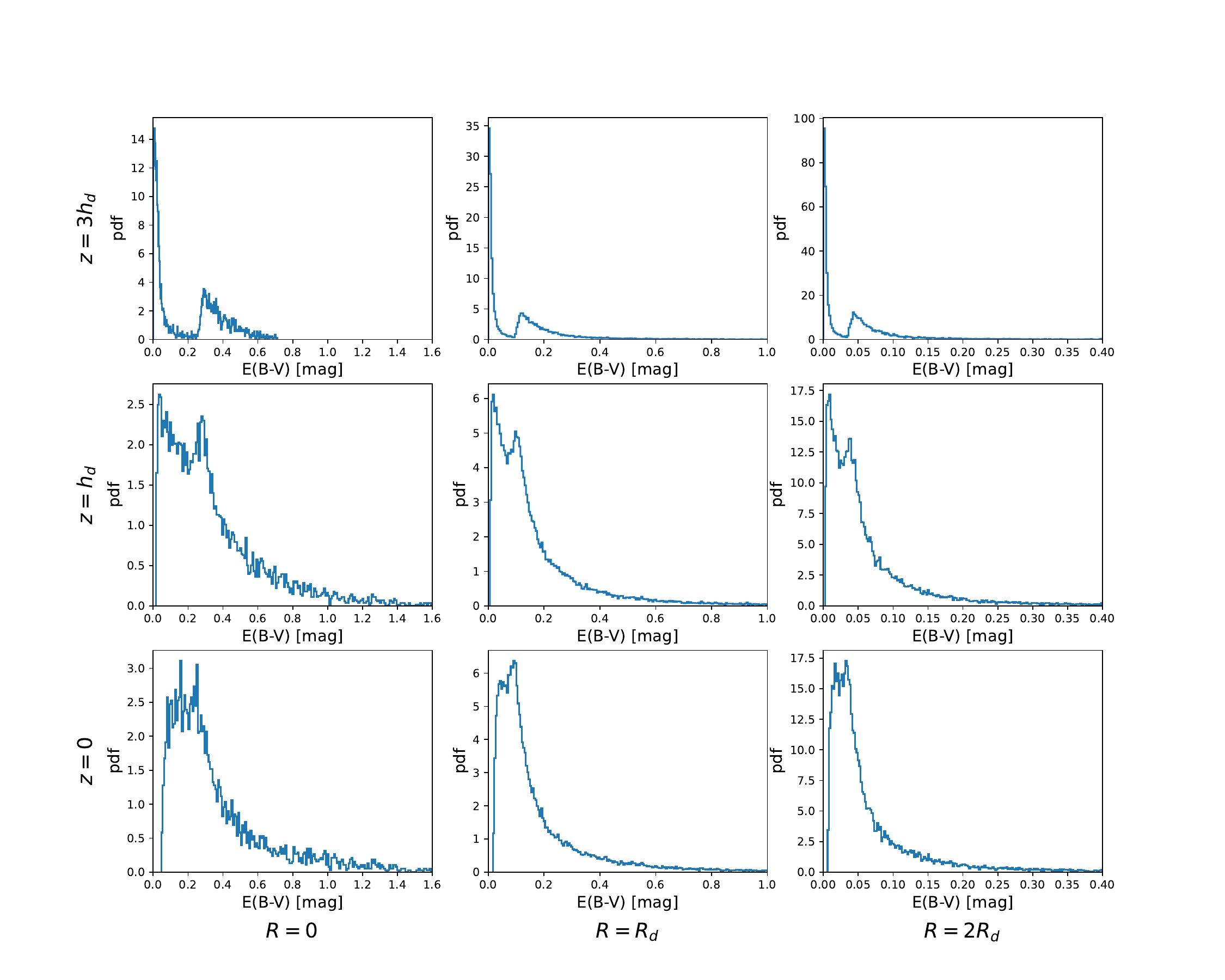}
    \caption{Distributions of host-galaxy interstellar reddening $E(B-V)$ for type Ia supernovae originating from 9 specific locations within the dust disc of a disc-dominated ($B/T=0$) late-type galaxy. The locations are distributed on a grid with (cylindrical) radii $R$ running across the columns and disc heights $z$ across the rows. Galaxy-observer orientations are assumed to be random and parameters of the dust disc are listed in Table~\ref{table:disky_setups}. Due to asymmetry between disc-supernova-observer and supernova-disc-observer orientations, the local distributions are bimodal for layers above the disc plane. The reddening distributions peak at $E(B-V)>0$ and vanish at $E(B-V)=0$ for $z\lesssim h_{\rm d}$.}
    \label{fig:BT0_Rzdiagram}
\end{figure*}

We derive the optical radius $R_{25}$ (the radius where the optical brightness in $B$-band is 25~${\rm mag}\,{\rm arcsec}^{-2}$) from the assumed stellar mass using an empirical scaling relation derived from observations from the Sloan Digital Sky Survey Data Release 15 (SDSS DR15) MaNGA survey \citep{Bundy2015,Pilyugin2021}. The optical radius is then used to estimate the scale parameters of the supernova disc adopting respectively $R_{\star}/R_{\rm 25}$ 
from \citet{casasola2017radial} and $h_{\rm Ia}/R_{25}$ measured directly from type Ia supernova vertical positions in edge-on galaxies \citep{hakobyan2017supernovae}. 
The effective radius is given by the ratio $R_{\rm e}/R_{\star}=0.2$ determined from fitting bulge and disc components to the light distribution in galaxies of different morphological types \citep{Laurikainen2010,Graham2008}.

For the interstellar dust distribution we assume an exponential disc model, i.e.
\begin{equation}
    \rho_{\rm disc,d}=\rho_{\rm 0} \exp\Big(-\frac{R}{R_{\rm d}}-\frac{|z|}{h_{\rm d}}\Big),
\end{equation}
where $\rho_{\rm 0}$ is a normalisation term derived from the total dust mass. We compute the total dust mass using the dust-to-stellar mass ratio $\log_{10}(M_{\rm d}/M_{\star}) = -3.1$, which is an average value measured in late-type galaxies with stellar masses of $10^{10}M_{\odot}$ \citep{Cortese2012}. The dust disc is assumed to be 1.8 times thinner than the stellar disc, 
i.e. $h_{\star}/h_{\rm d}\approx h_{\rm Ia}/h_{\rm d}=1.8$, and its scalelength is 1.4 times larger than the corresponding length of the stellar disc, as measured from observations
of edge-on late-type galaxies \citep{Xilouris1999,Galliano2018}. The obtained spatial scales of the dust disc are consistent with 
independent estimates derived from observational scaling relations involving the dust mass \citep{mosenkov2022distribution}. As suggested by observations, we assume that the dust disc extends twice as far as the light \citep{smith2016far}, where the maximum radius for the latter is given by the standard optical radius $R_{25}$. We also truncate the dust distribution at $6h_{\rm d}$ height above the disc plane.

\subsection{Lenticular galaxy (S0)}

We model S0-type galaxy as a transitional class between disc- and bulge-dominated galaxies. We use the above-described model devised for 
late-type galaxy with the bulge-to-total mass ratio $B/T=0.5$ which is typical value measured in lenticular galaxies \citep{Weinzirl2009,Laurikainen2007,Pastrav2021} and stellar mass $M_{\star}=10^{10.5}M_{\odot}$. Based on observational constraints, we assume that the total dust-to-stellar mass ratio is $\log_{10}(M_{\rm d}/M_{\star})=-4.4$ \citep{smith2012herschel,Alighieri2013}, which is $1.3$ dex lower than the corresponding ratio for late-type galaxies.

\subsection{Elliptical galaxy (E)}

Like in other cases, we assume that type Ia supernovae trace the stellar distribution. We model the stellar distribution in elliptical galaxies using the S\'ersic model with $n=4$. The effective radius of the S\'ersic model is derived from the observational scaling relation based on the stellar mass \citep{hyde2009curvature}. We adopt $10^{10.5}M_{\odot}$ as a representative stellar mass for this class of galaxies. We truncate the supernova distribution at $8R_{\rm e}$.

We compute the total dust mass using an average dust-to-stellar mass ratio $\log_{10}(M_{\rm d}/M_{\star})=-5.8$ measured in elliptical galaxies \citep{smith2012herschel,Alighieri2013}. 
The actual ratios can be higher than the adopted average due to evolutionary effects. Observations 
show that the dust content in early-type galaxies undergoes evolution driven by dust destruction processes \citep{Michalowski2019,Lesniewska2023}. Consequently, the dust-to-stellar mass ratio tends to be higher in dynamically younger galaxies. The highest dust-to-stellar mass ratios are comparable to those in late-type galaxies, i.e. $\log_{10}(M_{\rm d}/M_{\star})=-3$, and they are found in about 5--10 per cent of early-type galaxies \citep{Rowlands2012,Lesniewska2023}. In section~\ref{sect:disc}, we discuss implications of a scenario in which all early-type host galaxies of type Ia supernovae are maximally rich in dust.

Interstellar dust in early-type galaxies may exhibit complex spatial distributions with features often visible in optical light, such as lanes and patches. However, the majority of dust observed in infrared is thought to occur as a diffuse interstellar component \citep{Goudfrooij1995}. Given that infrared observations provide rather poor constraints on its spatial distribution, we assume that the interstellar dust traces the stellar mass.

\section{Results}
\label{sect:res}

\subsection{Reddening for different layers of dust disc}
\label{sect:layers}

\begin{figure*}[htbp]
    \centering
    \includegraphics[width=1\linewidth]{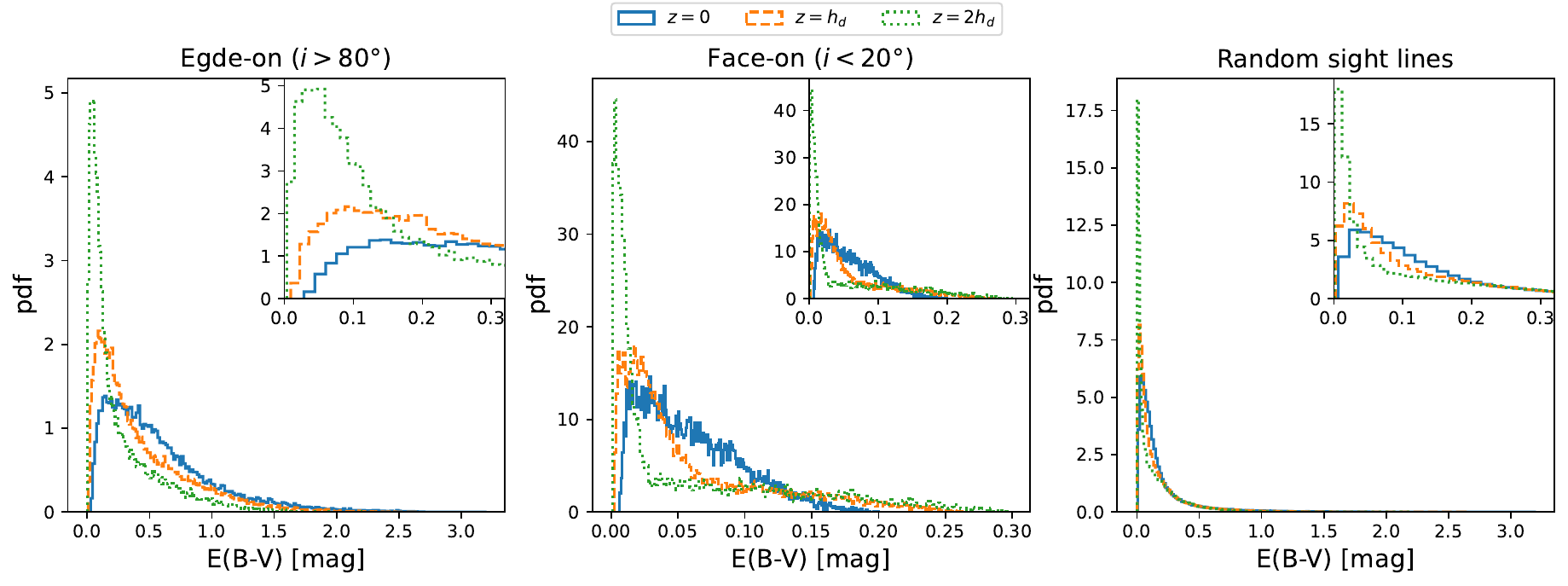}
    \caption{Distributions of host interstellar reddening $E(B-V)$ for type Ia supernovae originating from different layers in the dust disc of a late-type ($B/T=0$) galaxy, with vertical positions 
    $z=\{0,h_{\rm d}, 2h_{\rm d}\}$. The distributions are computed for edge-on (left panel), face-on (middle panel) and random (right panel) orientations of the galaxy disc with respect to an observer. Parameters of the dust disc are summarised in Table~\ref{table:disky_setups}. The results demonstrate that the distribution peak (the most probable reddening) depends strongly on supernova vertical position in the disc: the distributions are unimodal with maximum at $E(B-V)>0$ for $z\lesssim h_{\rm d}$ and monotonically decreasing, with maximum at $E(B-V)=0$ for $z>h_{\rm d}$.}   
    \label{fig:BT0_diffz_angleplot}
\end{figure*}

The reddening distribution depends both on the spatial distribution of dust and supernovae. The result of modelling is particularly sensitive to 
possible differences between the two distributions. Before we consider the distributions resulting from averaging over supernova positions in the host galaxy, it is instructive to begin with computing the expected reddening for several characteristic locations in the host galaxy. We use the model of a disc-dominated galaxy ($B/T=0$) from Table~\ref{table:disky_setups}.

Figure~\ref{fig:BT0_Rzdiagram} shows the expected reddening distribution for objects originating from 9 fixed positions in the dust disc. The positions are given by a grid with (cylindrical) radial coordinates $R=\{0,R_{\rm d},2R_{\rm d}\}$ and disc heights $z=\{0,h_{\rm d},3h_{\rm d}\}$, and the reddening distributions are generated for random sight lines. Comparing different panels of Figure~\ref{fig:BT0_Rzdiagram} we can see that the reddening distribution depends strongly on the place of supernova origin. For all locations 
above the disc plane, the distributions exhibit a bimodality which results from asymmetry of supernova-disc orientations with respect to an observer. 
The distribution peaks become gradually more asymmetric at larger heights above the disc plane where a smaller fraction of random sight lines 
can cross the dust disc and contribute to one of the distribution's peaks. The distributions are unimodal for supernovae located well within the dust disc at heights $z<h_{\rm d}$. In these cases, the reddening distributions reach maximum at $E(B-V)>0$ and vanish at $E(B-V)=0$.

Figure~\ref{fig:BT0_diffz_angleplot} shows the effect of averaging the local distributions shown in Figure~\ref{fig:BT0_Rzdiagram} over supernova radial positions. The panels show the reddening distributions for different disc orientations with respect to an observer. It is apparent that averaging over 
supernova radial positions erase bimodal features in the local distributions shown in Figure~\ref{fig:BT0_diffz_angleplot}. For supernovae originating from layers well within the dust disc ($z\lesssim h_{\rm d}$), the reddening distributions persist to peak at $E(B-V)>0$. The distributions become monotonically decreasing functions (maximum at $E(B-V)=0$) 
as soon as the supernova vertical position are comparable to the stellar disc scalelength $h_{\star}\approx h_{\rm Ia}\approx 2h_{\rm d}$. This transition appears to be the most pronounced for face-on configurations.

The distributions shown in Figure~\ref{fig:BT0_diffz_angleplot} may have relevance for modelling type Ia supernova observations. Although vertical positions within the stellar disc cannot be measured directly (unless the host galaxy is observed edge-on), we may expect a vertical stratification dictated by the progenitor age. Type Ia supernovae with short or long delay times are expected to trace respectively star formation regions or the old stellar population \citep{Mannucci2005,Scannapieco2005}. The resulting supernova populations would naturally exhibit different degrees of embedment in the dust disc (with the young population found preferentially in deeper layers of the dust disc) and thus different distributions of interstellar reddening.

\subsection{Reddening in late-type galaxies}

\begin{figure*}[htbp]
    \centering
    \includegraphics[width=1\linewidth]{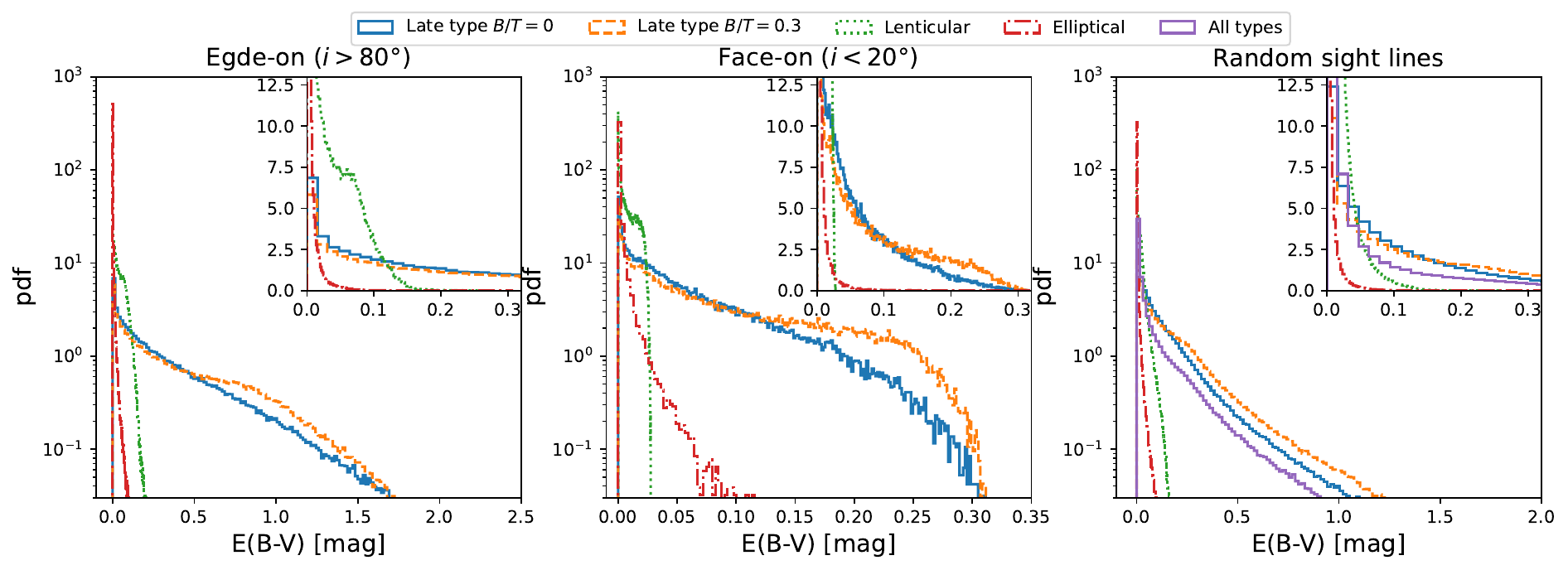}
    \caption{Distributions of host-galaxy interstellar reddening $E(B-V)$ for type Ia supernovae originating from host galaxies of four representative morphological types: late type with $B/T=0$ or $B/T=0.3$, S0 type and classical elliptical (for model parameters see Tables~\ref{table:disky_setups}-\ref{table:mprph_setups}). The distributions are computed for edge-on (left panel), face-on (middle panel) and random (right panel) orientations (no difference for elliptical galaxies which are assumed to be spherically symmetric). 
    The purple curve on the right panel shows the result of averaging type Ia supernova rates over host-galaxy morphological types. 
    All distributions peak at $E(B-V)\approx 0$. The apparent differences between the late- and early-type host galaxies are primarily driven by different dust-to-stellar mass ratios measured from infrared observations. The distributions shown in Figure~\ref{fig:BT0_diffz_angleplot} are strongly modified by averaging over supernova position in the host galaxy.
}
    \label{fig:morph_angleplot}
\end{figure*}

Figure~\ref{fig:morph_angleplot} shows the reddening distribution resulting from averaging over supernova positions in our model of late-type host galaxy ($B/T=0$ or $B/T=0.3$). It is 
apparent that the distribution tail depends strongly on the galaxy inclination. For face-on galaxies ($i<20^{\circ}$), the distributions appear to be cut off at $E(B-V)\approx 0.3\,{\rm mag}$ which marks the maximum dust column density expected for $R\ll R_{\rm d}$ and $z\ll h_{\rm d}$. On the other hand, the distributions for edge-on 
orientations ($i>80^{\circ}$) exhibit long tails with reddening reaching $E(B-V)\gtrsim 1\,{\rm mag}$ corresponding to supernovae well embedded within the dust disc. The distributions converge to an exponentially decaying probability density at $E(B-V)\gtrsim 0.1\,{\rm mag}$. Averaging over random sight lines results in distributions for which exponential dependence on $E(B-V)$ can be extended to a wider range of values. However, a clear deviation from the exponential distribution can be seen for the smallest (and most probable) reddening. Here, the simulations predict a higher probability of incidents with zero interstellar reddening than the exponential model.

As shown in Figure~\ref{fig:morph_angleplot}, the effect of bulge on the reddening distributions is noticeable, but relatively small. 
Supernovae originating from the bulge component have more elevated positions than those in the stellar disc, but relatively shorter distances from the vertical centre axis perpendicular to the disc plane. This explains the enhanced 
probability of high ($E(B-V)\gtrsim 0.2\,{\rm mag}$) and very small ($E(B-V)\approx 0$) reddening for face-on galaxies. More frequent occurrence of higher reddening is also visible for edge-on galaxies. This can be linked to a higher concentration of supernovae with vertical positions coinciding with the disc plane.

The apparent differences between purely discy and bulged galaxies depend quite strongly on the effective radius of the bulge relative to the stellar disc radial and vertical scale. The bulge effective radius adopted in our simulations and based on observational properties of disc-like galaxies is about 5 times smaller than the stellar disc scalelength and comparable to its scaleheight. One can imagine that modifying the proportions between the scales of the bulge and the stellar disc can substantially change our results. For example, it is possible to increase the probability of sight lines with zero dust column densities ($E(B-V)\approx 0$) without any limit by increasing the bulge effective radius and the bulge-to-disc ratio of supernova rates. However, the resulting galaxy models would eventually involve unrealistic bulges which are not represented in observations.

\begin{figure}[htbp]
    \centering
    \includegraphics[width=1\linewidth]{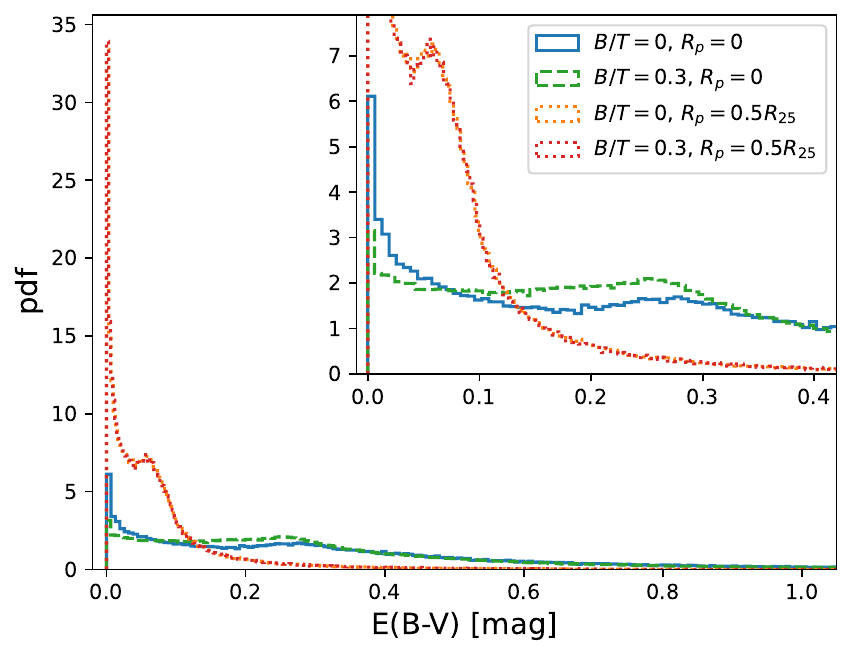}
    \caption{Distributions of host-galaxy interstellar reddening $E(B-V)$ for type Ia supernovae at two different projected distance from the centre of a late-type ($B/T=0$ or $B/T=0.3$) host galaxy: $R_{\rm p} = 0$ and $R_{\rm p} = 0.5R_{25}$. The figure demonstrates a strong dependence of the conditional reddening on supernova location in the host galaxy.}
    \label{fig:Rp_plot}
\end{figure}

Table~\ref{table:ebv} provides summary statistics for the simulated reddening distributions in randomly distributed sight lines. We also compute analogous summary statistics for two fixed projected radii $R_{\rm p}=\{0,0.5R_{25}\}$ and show the corresponding distributions in Figure~\ref{fig:Rp_plot}. This is motivated
by observations showing different degrees of scatter in supernova Hubble residuals across
the projected distance from the host galaxy centre \citep{Uddin2024,Toy2025}. These observations are typically interpreted as an effect of differences between the reddening distribution in the inner and outer parts of supernova host galaxies. Figure~\ref{fig:Rp_plot} confirms this scenario.

Although our simulations are obtained for a fixed stellar mass, the expected 
reddening can be estimated using the assumed scaling relations between scale parameters of the discs and $R_{25}$. Observationally motivated scalability of the dust and stellar disc sizes with $R_{25}$ 
implies that reddening $E(B-V)$ expected at any stellar mass can be derived from the following relation:
\begin{equation}
E(B-V)\propto \frac{M_{\rm d}}{R_{25}^{2}}=\frac{(M_{\rm d}/M_{\star})M_{\star}}{R_{25}^{2}}\propto M_{\star}^{-0.2},
\label{eq:tau_Mstar_eq}
\end{equation}
where we assume $R_{25}\propto M_{\star}^{0.26}$ derived from observations \citep{Pilyugin2021} 
and $M_{\rm d}/M_{\star}\propto M_{\star}^{-0.7}$ approximating the dust-to-stellar mass ratios measured 
as a function of the stellar mass \citep{Cortese2012}. The relation can be applied to any $E(B-V)$ estimate including 
percentiles or moments of the simulated distributions. For $10^{11}M_{\odot}$ and $10^{9}M_{\odot}$ as the upper and lower stellar mass limits for type Ia supernova host galaxies, the difference in reddening is about $0.4$~dex. 
This is comparable to the expected scatter of $0.5$~dex driven by the scatter in the observed dust-stellar mass relations \citep{Cortese2012}.

\begin{table*}
\caption{Summary statistics for interstellar host-galaxy reddening $E(B-V)$ obtained from simulating dust columns densities in random sight lines to type Ia supernovae.
}            
\label{table:ebv}     
\centering          
\begin{tabular}{l|cccc|ccc|c|c|c}     
\hline     
    & \multicolumn{4}{c}{LT (B/T=0)} &      \multicolumn{3}{c}{LT (B/T=0.3)}  & S0 & E    & All types             \\
    & any $R_{\rm p}$ & $R_{\rm p}=0$ & $R_{\rm p}=0.5R_{25}$ & $z = 0$ & any $R_{\rm p}$ & $R_{\rm p}=0$ & $R_{\rm p}=0.5R_{25}$ & any $R_{\rm p}$ & any $R_{\rm p}$ & any $R_{\rm p}$\\
\hline
 $\langle E(B-V)\rangle$ & 0.15 & 0.35 & 0.08 & 0.20 & 0.19 & 0.33 & 0.08 & 0.019 & 0.002   & 0.09 \\   
 $E(B-V)_{25\%}$         & 0.02 & 0.10 & 0.02 & 0.06 & 0.03 & 0.13 & 0.02 & 0.004 & 0.00004 & 0.001 \\  
 $E(B-V)_{50\%}$         & 0.08 & 0.27 & 0.05 & 0.11 & 0.11 & 0.26 & 0.5 & 0.013 & 0.0002  & 0.019 \\  
 $E(B-V)_{75\%}$         & 0.19 & 0.48 & 0.09 & 0.22 & 0.26 & 0.44 & 0.09 & 0.026 & 0.0013  & 0.10 \\  
 $E(B-V)_{95\%}$         & 0.56 & 1.05 & 0.26 & 0.71 & 0.67 & 0.95 & 0.26 & 0.064 & 0.010   & 0.43 \\  
 $\gamma$                & 0.62 & 0.91 & 0.70 & 1.03 & 0.66 & 1.17 & 0.70 & 0.71 & 0.29     & 0.30 \\  
    \hline                
\end{tabular}
\tablefoot{The results are based on modelling the dust and type Ia supernova distributions in galaxies representing four morphological types (see details in Tables~\ref{table:disky_setups}-\ref{table:mprph_setups}). 
For late-type galaxies the table  provides  also summary statistics conditioned on projected distances $R_{\rm p}$ between supernovae and the host-galaxy centre (two cases with $R_{\rm p}=0$ and $R_{\rm p}=0.5R_{25}$) or supernovae originating from the $z=0$ plane of the disc-dominated ($B/T = 0$) galaxy. $\langle \rangle$ and "$_{\rm k}$" denote the mean and $k$-th percentile, $\gamma$ is the best-fit parameter of the $\gamma$-distribution model fitted to the simulated data.}
\end{table*}

\subsection{Reddening across morphological types}

Figure~\ref{fig:morph_angleplot} compares the reddening distributions simulated for four morphological types of supernova host galaxies. The scale of simulated interstellar reddening spans two orders of magnitude, with mean reddening of $0.15$~mag in late-type galaxies, $0.02$~mag in S0-type hosts and $0.002$~mag in ellipticals (see Table~\ref{table:ebv} for the complete summary statistics). This is primarily driven by a large dynamic range of dust-to-stellar mass ratios across galaxy morphological types. The assumed dust and supernova spatial distributions have mainly impact on the distribution shape. Figure~\ref{fig:morph_angleplot} shows that the reddening distribution obtained for elliptical galaxies exhibits the fastest growth at small reddening.

Since many observational estimates of interstellar reddening in supernova host galaxies are derived probabilistically from entire samples without distinction between supernova host-galaxy types, it is instructive to compute the expected distribution reflecting the actual supernova rates across galaxy morphological types. Based on morphological type estimates from HyperLeda\footnote{http://leda.univ-lyon1.fr} \citep{Makarov2014} obtained for host galaxies of type Ia supernovae from the SuperCal compilation \citep{Scolnic2015} at redshifts $z<0.15$ we find 25 per cent of the supernovae in elliptical galaxies (morphological types, as introduced by \citet{DeVa1974}, $t\in[-5,-3]$), 25 per cent in S0-type galaxies ($t\in(-3,1]$) and 50 per cent in late-type galaxies ($t\in(1,7]$), in fair agreement with other available studies \citep[see e.g.][]{Pruzhinskaya2020}. For the late-type type, we assume equal split between bulged and disc-dominated galaxies. Using these fractions as the weights for sampling supernovae across our host-galaxy models we obtain the distribution which shows our predictions marginalised over supernova host-galaxy morphological types. Figure~\ref{fig:morph_angleplot} shows the resulting distribution and the last column of Table~\ref{table:ebv} lists the quantiles. The obtained average distribution represents low-redshift observations. It is expected to be modified at high redshifts due to evolutionary processes changing the relative fractions of host-galaxy morphological types and the dust content.

\subsection{Exponential model}

\begin{figure*}[htbp]
    \centering
    \includegraphics[width=0.85\linewidth]{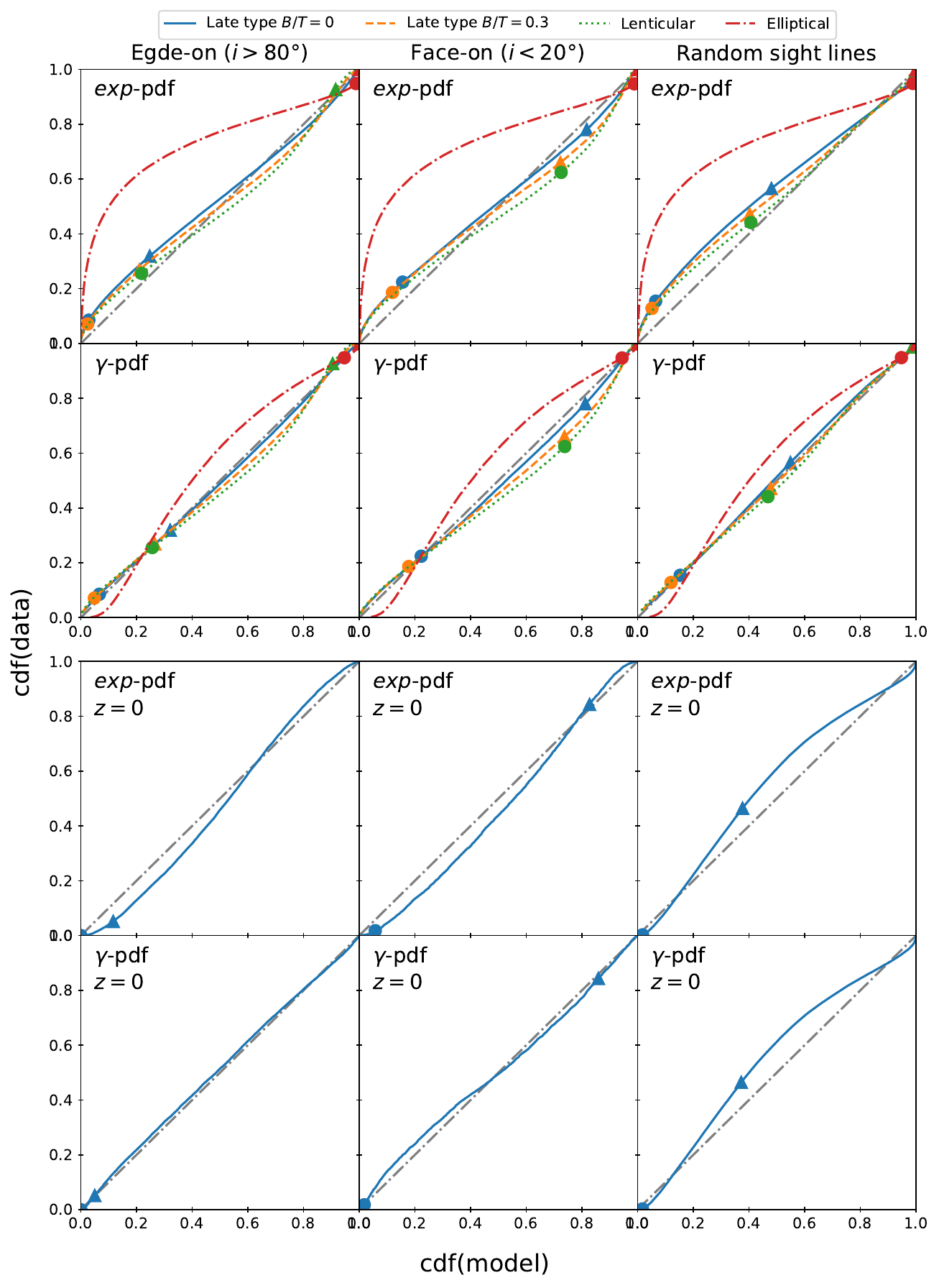}
    \caption{Comparison between the cumulative distribution of simulated host-galaxy reddening $E(B-V)$ (y-axis) and its approximation (x-axis) given by the best-fit exponential (exp-pdf) or $\gamma$-distribution ($\gamma$-pdf) model. Each panel in the upper half shows results for four morphological types of simulated host galaxies: late type (with $B/T=0$ and $B/T=0.3$), S0 type and early type. The columns show three cases of different galaxy-observer orientations: edge-on (left), face-on (middle) and random (right). The bottom panels show analogous comparison for supernovae originating from the $z=0$ plane of the dust disc in the late-type host-galaxy model ($B/T=0$). The symbols denote two characteristic scales of reddening with $E(B-V)=0.01$~mag ($\bullet$) and $E(B-V)=0.1$~mag ($\blacktriangle$), as given by the simulated distributions. The exponential model underestimates the probability of zero reddening, especially in the early-type host galaxies. The simulated data are better reproduced by the $\gamma$-distribution. 
    For type Ia supernovae originating from the proximity of the $z=0$ disc plane, the exponential model overpredicts the supernova rate with $E(B-V)\approx 0$.}
    \label{fig:cdf_plots}
\end{figure*}

Figure~\ref{fig:morph_angleplot} shows that an exponential dependence of the simulated distributions on reddening is only approximate. In order to quantify 
the accuracy of the exponential model in reproducing the simulated data, we compare the cumulative distributions of the model and the simulated host-galaxy reddening. The model is given by
\begin{equation}
    p(E(B-V))=\frac{1}{\tau}\exp\Big(-\frac{E(B-V)}{\tau}\Big),
\end{equation}
where the scale parameter $\tau$ is by construction equal to the mean reddening $\langle E(B-V)\rangle$, which is computed directly from the simulated data. The results are shown in Figure~\ref{fig:cdf_plots} for all four morphological types considered in our study and for three cases of host-galaxy orientations (top panels). In the lower panels, we also show analogous test for supernovae originating from the $z=0$ plane of the dust disc in late-type galaxies.

Figure~\ref{fig:cdf_plots} demonstrates that the simulated distributions depart from their exponential approximations. In general, the simulations generate a higher fraction of sight lines with zero reddening ($E(B-V)\approx 0$) with respect to the exponential model. This property is particularly well visible for early-type galaxies, but it is more subtle for late-type galaxies. We find that a better match to the simulated distributions can 
be obtained using the $\gamma$-distribution given by
\begin{equation}
p(E(B-V))=\frac{1}{\tau\Gamma(\gamma)}\exp\Big(-\frac{E(B-V)}{\tau}\Big)\Big(\frac{E(B-V)}{\tau}\Big)^{\gamma-1},
\label{pdf_gamma}
\end{equation}
which is the simplest generalisation of the exponential model (recovered for $\gamma=1$), where $\Gamma(\gamma)$ is the gamma function. Fitting the model to the simulated reddening we find strong evidence for $\gamma<1$ for all host-galaxy morphological types. This signifies more pronounced peaks of the distributions at $E(B-V)\approx 0$ than the exponential model with the same scale parameter. The lowest value of $\gamma$ is obtained for the elliptical host galaxy. The best-fit values are listed in Table~\ref{table:ebv}.

Population of type Ia supernovae originating from the $z=0$ plane (which can be thought of as a close tracer of prompt supernovae) exhibits a different pattern of differences with respect to the approximating exponential model. Here, the exponential model overpredicts the supernova rate with $E(B-V)\approx 0$ and underpredicts those with large reddening. We find that the simulated data are mildly better approximated by the $\gamma$-distribution model with $\gamma\gtrsim 1$ (see Table~\ref{table:ebv}).

\subsection{Supernova colour distribution}

\begin{figure*}    
\centering
    \includegraphics[width=1\linewidth]{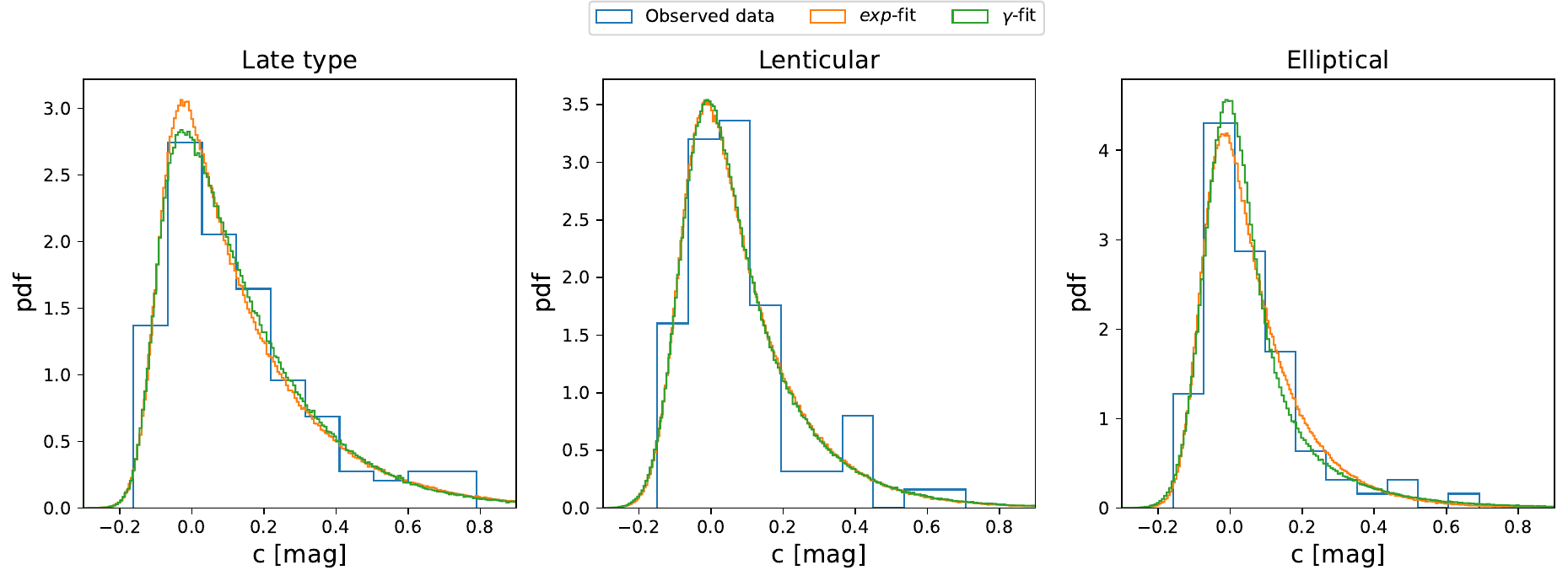}
    \caption{Distribution of type Ia supernova colour parameter $c$ in late-type (left panel), S0-type (middle panel) and elliptical (right panel) galaxies. The coarse-bin histogram show the measurements from the ZTF volume-limited sample of normal type Ia supernovae and fine-bin histograms are best-fit models assuming a Gaussian prior distribution for intrinsic colours and a $\gamma$-distribution (or exponential) model for reddening $E(B-V)$.}
    \label{c_pdf}
\end{figure*}

The main observational manifestation of dust reddening lies in long tails of the observed supernova colour distributions at red colours. Here we use measurements of the apparent $B-V$ colour (at the peak of the B-band light curve) given by the colour parameter $c$ of the SALT light curve fitter \citep{Guy2005} to obtain constraints on free parameters of the expected reddening distribution given by eq.~(\ref{pdf_gamma}), and compare them to the predictions of the dust model. We model the observed colour as the sum of a priori unknown time-independent component of intrinsic colour $c_{\rm int}$ and reddening $E(B-V)$, i.e.
\begin{equation}
    c = c_{\rm int}+E(B-V).
\end{equation}
We use colour parameter measurements from the volume-limited ($z<0.06$) sample of normal type Ia from the Zwicky Transient Factory (ZTF) DR2 \citep{Rigault2025}. We 
include supernovae with the recommended high quality fits given by a set of criteria outlined in \citet{Rigault2025}. We split the sample into three groups of host-galaxy morphological types corresponding to the three morphological classes adopted for our dust modelling. We use a continuous morphological type $t$ introduced by \citet{DeVa1974} and its estimates from the HyperLeda database \citep{Makarov2014}\footnote{ http://leda.univ- lyon1.fr}. With the morphological type estimates available for 38 per cent of the supernova host galaxies, we find 74 supernovae in elliptical ($t\in[-5,-3]$), 73 in S0-type ($t\in(-3,1]$) and 153 late-type ($t\in(1,7]$) galaxies. Figure~\ref{c_pdf} shows the distribution of supernova colour parameter $c$ in the three groups of host-galaxy morphological types. 
The mean logarithmic stellar masses computed from the ZTF host-galaxy data are: $10.21$ (late type), $10.51$ (S0 type) and $10.50$ (early type). They match closely the fiducial stellar masses assumed in our dust model.

We fit the distribution of the supernova colour parameter assuming a model 
given by a convolution of a Gaussian prior distribution for $c_{\rm int}$ with the prior reddening distribution given by eq.~(\ref{pdf_gamma}) and a zero-mean Gaussian accounting for measurement uncertainty per supernova. Although gaussianity of the intrinsic colour distribution is commonly taken for granted, this assumption has never been verified on theoretical and observational grounds making it the main source of potential systematic errors in derived properties of the reddening distribution.

\begin{figure}    
\centering
    \includegraphics[width=1\linewidth]{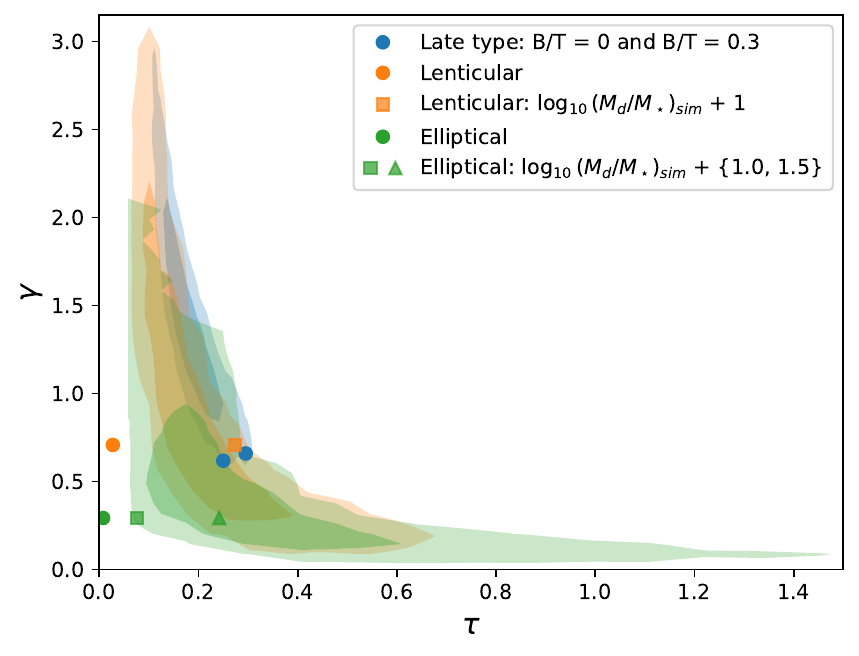}
    \caption{Parameters of the reddening distribution (eq.~\ref{pdf_gamma}) obtained from fitting the observed supernova colours conditioned on gaussianity of the intrinsic colour distribution. The results shown as 
    68 and 95 per cent credibility contours are compared to the predictions of the dust model from this study. The apparent discrepancy between the observational constraints and the dust model for early-type host galaxies implies that either the observed red colours are primarily driven by a larger variation in supernova intrinsic colours or early-type host galaxies are unusually rich in interstellar dust. The square and triangle symbols show the effect of increasing the dust-to-stellar mass ratio by 1.0~dex and 1.5~dex.}
    \label{gamma_tau}
\end{figure}

Figure~\ref{gamma_tau} shows constraints on free parameters of the prior reddening distribution obtained in each host-galaxy morphological class assuming independent prior distributions of the intrinsic colours. The credibility contours were obtained by integrating the posterior probability distributions using a \textit{Monte Carlo Markov Chain} method implemented in the \textit{emcee} code \citep{emcee}. The adopted likelihood was given by the product of the model distribution evaluated at supernova colour parameter measurements. The best-fit model distributions of the supernova colour parameter are compared to the data in Figure~\ref{c_pdf}. The distributions are computed by means of sampling from the prior distributions with best-fit hyperparameters and from the actual distribution of measurement uncertainties.

As shown in Figure~\ref{gamma_tau}, we find only a mild $\sim 2\sigma$ difference between the observationally constrained reddening distribution and the dust model for late-type host galaxies. On the other hand, the observations point to a significantly larger scale parameter of the reddening distribution in early-type galaxies than the prediction of our dust model. This implies that the apparent excess of red supernovae in early-type host galaxies ($c\gtrsim 0.2$~mag in Figure~\ref{c_pdf}) cannot be explained by reddening based on a typical interstellar dust content in these types of host galaxies. There are two possible solutions to this discrepancy. The first possibility is to assume that the excess of apparent red colours is driven by the variation in intrinsic colours. In this scenario, the distribution of intrinsic colours is not Gaussian and it exhibits a longer tail at red colours. Another possibility is to assume that early-type host galaxies are richer in dust than average dust-to-stellar mass ratios. As shown in Figure~\ref{gamma_tau}, this would require a 1.5~dex (1.0~dex) larger dust-to-stellar mass ratios for elliptical (S0-type) galaxies. These relatively high dust-to-stellar mass ratio are observed only in a fraction of early-type galaxies.
   
\section{Discussion}
\label{sect:disc}

Deriving cosmological distances from type Ia supernova observations depend on model-dependent extinction corrections conditioned on apparent supernova colours. The standard assumption used commonly for modelling type Ia supernova colours is that the host-galaxy reddening distribution is universal across host-galaxy morphological types and well approximated by the exponential model. However, this cannot be reconciled with the results presented in our study. Firstly, our simulations of host-galaxy reddening demonstrate substantial differences between the interstellar reddening distributions across host-galaxy morphological types. Secondly, the simulated distributions deviate from the exponential model and predict a more frequent occurrence of sight lines with zero reddening than the corresponding best-fit exponential models. Departure from the exponential model appears to be particularly strong for early-type host galaxies and less prominent for late-type host galaxies. We find that the simulated reddening distributions are better reproduced by a $\gamma$-distribution with $\gamma<1$.  The model predicts higher probability at $E(B-V)=0$ than the corresponding exponential distribution for which $\gamma=1$.  Similar values of the shape parameter $\gamma$ were found by \citet{Ward2023} for a supernova sample including all highly reddened supernovae.

The obtained mean reddening for late-type galaxies, i.e. $\langle E(B-V)\rangle=0.15$~mag, agrees very well with observational constraints from modelling supernova colours,  
ranging between $\langle E(B-V)\rangle\approx 0.1$ mag and  $\langle E(B-V)\rangle\approx 0.15$ mag \citep[see e.g.][]{Brout2021,Ginolin2025,Wojtak2023,Ward2023}. We also predict a similar rate of the most reddened supernovae: 5 per cent of the most reddened supernovae in our simulations have $E(B-V)>0.5$~mag, while observations yield $E(B-V)\gtrsim 0.4$~mag based on redshift $z<0.2$ supernovae from the Pantheon+ compilation without any cuts in colour \citep{Rose2022} and assuming mean intrinsic colour of $-0.08$~mag \citep{Popovic2021}.

Our results suggest that the interstellar reddening in early-type galaxies should not exceed $E(B-V)\approx 0.05$ mag due to a substantially lower dust-to-stellar mass ratio found on average in these galaxies. This appears to disagree with the fact that the observed supernova colours do not show a clear suppression of the red tail for early-type host galaxies \citep{Ginolin2025,Larison2024}. In fact, \citet{Ginolin2025} found only $0.06$~mag difference between the mean reddening in supernova host galaxies rich in dust and those selected as dustless. This apparent discrepancy between the expected and observationally estimated reddening in early-type galaxies, 
which is shown explicitly in Figure~\ref{gamma_tau}, can be solved by assuming that elliptical galaxies producing type Ia supernova are not typical and they are as rich in dust as late-type counterparts. The challenge here is that only about 5--10 per cent of early-type galaxies have dust masses comparable to late-type analogs \citep{Rowlands2012,Lesniewska2023}. Another option is that the observed red colours of type Ia supernovae originating from early-type galaxies (or low-stretch supernovae) are not due to host-galaxy interstellar reddening but to a larger extent due to variation in intrinsic colours \citep[see also][]{Burns2014,Gonzalez2011}. This scenario can potentially help alleviate the problem of oddly low estimates of the total-to-selective extinction coefficients with $R_{\rm B}\approx 3$ \citep[for a compilation of measurements from the literature see][]{Thorp2024}. These estimates are typically obtained for low-stretch type Ia supernovae \citep{Wojtak2025} or supernovae originating from massive galaxies \citep{Brout2021}, where fits are strongly driven by supernova originating from elliptical galaxies. Their values coincide closely with the coefficient quantifying the dependence of supernova peak magnitude on its intrinsic colour \citep{Brout2021,Wojtak2025}. This coincidence suggests that what is modelled as two independent effects -- host-galaxy reddening and supernova intrinsic colours -- can be primarily driven solely by the former, and support the conjecture that highly reddened supernova in early-type host galaxies result from the variation in supernova intrinsic colours. Further tests 
of this scenario should involve Bayesian hierarchical modelling with a wider range of possible prior distributions of intrinsic colours for supernovae originating from early-type galaxies.

The simulated reddening in late-type host galaxies is quite sensitive to the distribution of supernova vertical positions in the disc. For type Ia supernovae that are more embedded in the dust disc, the reddening distributions tend to peak at nonzero values. This configuration may be preferred by a population of young type Ia supernovae originating from star forming environments which are more correlated with the dust disc and less extended in vertical positions than the stellar disc. The young supernova population is hypothesised as one of the two progenitor channels required to explain the observed distribution of delay times between the star formation and type Ia supernova explosions \citep{Mannucci2005,Scannapieco2005}. Relative excess of this population may be expected in galaxies with high recent star formation. 
Possible connection between recent star formation histories and the extent of supernova vertical positions in the disc is a testable scenario, especially with large surveys such as the Large Synoptic Survey Telescope (LSST) of the Rubin Observatory \citep{LSST2009}. Interestingly, several recent analyses of type Ia supernova data employing the $\gamma$-distribution, as a generalisation of the exponential model, for host-galaxy reddening show a preference for distributions peaking at 
nonzero $E(B-V)$ expected for young supernova populations \citep{Wojtak2023,Ward2023}.

Using type Ia supernova host galaxy-dependent reddening prior is also relevant for deriving unbiased cosmological distances. This becomes particularly relevant when distance estimates are propagated between type Ia supernovae originating from 
different samples of host galaxies. The determination of the Hubble constant based on distance calibration from Cepheids is a prime example: type Ia supernovae from the Hubble flow originate from galaxies of mixed morphological types, whereas the calibration galaxies with observable Cepheids are exclusively late-type galaxies. The leading Hubble constant measurement of this kind obtained by \citet{Riess2022} is based on the assumption that prior reddening distribution and derived extinction parameters are the same both in the calibration galaxies and the Hubble flow \citep{Popovic2021,Brout2021}. Direct effect of assuming a reddening prior distribution that is shared across different morphological types of supernova host galaxies is an underestimation of extinction corrections in the calibration supernovae (due to low effective extinction extrapolated from elliptical galaxies in the Hubble flow) which in turn results in an overestimation of the Hubble constant \citep{Wojtak2022,Wojtak2024,Wojtak2025}. An effective way to avoid this kind of bias is to separate reddening priors for early- and late-type host galaxies \citep{Wojtak2025} or to consider only one class of morphological types \citep{Newman2025}. Given the fact that the physical origin of reddened supernovae in early-type supernova is yet to be understood, these strategies are necessary to follow in order to minimise systematic errors in the Hubble constant estimation.

\begin{acknowledgements}
This work was supported by research grants (VIL16599,VIL54489) from VILLUM FONDEN. RW thanks Jo\~{a}o Duarte and Santiago Gonz\'{a}lez-Gait\'{a}n for inspiring discussions and comments on the paper. The authors thank the anonymous referee for useful comments that helped improve this work.
\end{acknowledgements}

\bibliographystyle{aa}
\bibliography{master}

\begin{thebibliography}{87}
\expandafter\ifx\csname natexlab\endcsname\relax\def\natexlab#1{#1}\fi

\bibitem[{{Adame} {et~al.}(2025){Adame}, {Aguilar}, {Ahlen}, {Alam},
  {Alexander}, {Alvarez}, {Alves}, {Anand}, {Andrade}, {Armengaud}, {Avila},
  {Aviles}, {Awan}, {Bahr-Kalus}, {Bailey}, {Baltay}, {Bault}, {Behera},
  {BenZvi}, {Bera}, {Beutler}, {Bianchi}, {Blake}, {Blum}, {Brieden},
  {Brodzeller}, {Brooks}, {Buckley-Geer}, {Burtin}, {Calderon}, {Canning},
  {Carnero Rosell}, {Cereskaite}, {Cervantes-Cota}, {Chabanier}, {Chaussidon},
  {Chaves-Montero}, {Chen}, {Chen}, {Claybaugh}, {Cole}, {Cuceu}, {Davis},
  {Dawson}, {de la Macorra}, {de Mattia}, {Deiosso}, {Dey}, {Dey}, {Ding},
  {Doel}, {Edelstein}, {Eftekharzadeh}, {Eisenstein}, {Elliott}, {Fagrelius},
  {Fanning}, {Ferraro}, {Ereza}, {Findlay}, {Flaugher}, {Font-Ribera},
  {Forero-S{\'a}nchez}, {Forero-Romero}, {Frenk}, {Garcia-Quintero},
  {Gazta{\~n}aga}, {Gil-Mar{\'\i}n}, {Gontcho a Gontcho}, {Gonzalez-Morales},
  {Gonzalez-Perez}, {Gordon}, {Green}, {Gruen}, {Gsponer}, {Gutierrez}, {Guy},
  {Hadzhiyska}, {Hahn}, {Hanif}, {Herrera-Alcantar}, {Honscheid}, {Howlett},
  {Huterer}, {Ir{\v{s}}i{\v{c}}}, {Ishak}, {Juneau}, {Kara{\c{c}}ayl{\i}},
  {Kehoe}, {Kent}, {Kirkby}, {Kremin}, {Krolewski}, {Lai}, {Lan}, {Landriau},
  {Lang}, {Lasker}, {Le Goff}, {Le Guillou}, {Leauthaud}, {Levi}, {Li},
  {Linder}, {Lodha}, {Magneville}, {Manera}, {Margala}, {Martini}, {Maus},
  {McDonald}, {Medina-Varela}, {Meisner}, {Mena-Fern{\'a}ndez}, {Miquel},
  {Moon}, {Moore}, {Moustakas}, {Mueller}, {Mu{\~n}oz-Guti{\'e}rrez}, {Myers},
  {Nadathur}, {Napolitano}, {Neveux}, {Newman}, {Nguyen}, {Nie}, {Niz},
  {Noriega}, {Padmanabhan}, {Paillas}, {Palanque-Delabrouille}, {Pan},
  {Penmetsa}, {Percival}, {Pieri}, {Pinon}, {Poppett}, {Porredon}, {Prada},
  {P{\'e}rez-Fern{\'a}ndez}, {P{\'e}rez-R{\`a}fols}, {Rabinowitz}, {Raichoor},
  {Ram{\'\i}rez-P{\'e}rez}, {Ramirez-Solano}, {Rashkovetskyi}, {Ravoux},
  {Rezaie}, {Rich}, {Rocher}, {Rockosi}, {Roe}, {Rosado-Marin}, {Ross},
  {Rossi}, {Ruggeri}, {Ruhlmann-Kleider}, {Samushia}, {Sanchez}, {Saulder},
  {Schlafly}, {Schlegel}, {Schubnell}, {Seo}, {Shafieloo}, {Sharples},
  {Silber}, {Slosar}, {Smith}, {Sprayberry}, {Tan}, {Tarl{\'e}}, {Taylor},
  {Trusov}, {Ure{\~n}a-L{\'o}pez}, {Vaisakh}, {Valcin}, {Valdes},
  {Vargas-Maga{\~n}a}, {Verde}, {Walther}, {Wang}, {Wang}, {Weaver},
  {Weaverdyck}, {Wechsler}, {Weinberg}, {White}, {Yu}, {Yu}, {Yuan},
  {Y{\`e}che}, {Zaborowski}, {Zarrouk}, {Zhang}, {Zhao}, {Zhao}, {Zhou}, \&
  {Zhuang}}]{Adame2025}
{Adame}, A.~G., {Aguilar}, J., {Ahlen}, S., {et~al.} 2025, \jcap, 2025, 021

\bibitem[{{Anderson} {et~al.}(2015){Anderson}, {James}, {F{\"o}rster},
  {Gonz{\'a}lez-Gait{\'a}n}, {Habergham}, {Hamuy}, \& {Lyman}}]{Andersen2015}
{Anderson}, J.~P., {James}, P.~A., {F{\"o}rster}, F., {et~al.} 2015, \mnras,
  448, 732

\bibitem[{{Barkhudaryan}(2023)}]{Barkhudaryan2023}
{Barkhudaryan}, L.~V. 2023, \mnras, 520, L21

\bibitem[{{Brout} \& {Scolnic}(2021)}]{Brout2021}
{Brout}, D. \& {Scolnic}, D. 2021, \apj, 909, 26

\bibitem[{{Brout} {et~al.}(2022){Brout}, {Scolnic}, {Popovic}, {Riess}, {Carr},
  {Zuntz}, {Kessler}, {Davis}, {Hinton}, {Jones}, {Kenworthy}, {Peterson},
  {Said}, {Taylor}, {Ali}, {Armstrong}, {Charvu}, {Dwomoh}, {Meldorf},
  {Palmese}, {Qu}, {Rose}, {Sanchez}, {Stubbs}, {Vincenzi}, {Wood}, {Brown},
  {Chen}, {Chambers}, {Coulter}, {Dai}, {Dimitriadis}, {Filippenko}, {Foley},
  {Jha}, {Kelsey}, {Kirshner}, {M{\"o}ller}, {Muir}, {Nadathur}, {Pan}, {Rest},
  {Rojas-Bravo}, {Sako}, {Siebert}, {Smith}, {Stahl}, \& {Wiseman}}]{Brout2022}
{Brout}, D., {Scolnic}, D., {Popovic}, B., {et~al.} 2022, \apj, 938, 110

\bibitem[{{Bundy} {et~al.}(2015){Bundy}, {Bershady}, {Law}, {Yan}, {Drory},
  {MacDonald}, {Wake}, {Cherinka}, {S{\'a}nchez-Gallego}, {Weijmans}, {Thomas},
  {Tremonti}, {Masters}, {Coccato}, {Diamond-Stanic}, {Arag{\'o}n-Salamanca},
  {Avila-Reese}, {Badenes}, {Falc{\'o}n-Barroso}, {Belfiore}, {Bizyaev},
  {Blanc}, {Bland-Hawthorn}, {Blanton}, {Brownstein}, {Byler}, {Cappellari},
  {Conroy}, {Dutton}, {Emsellem}, {Etherington}, {Frinchaboy}, {Fu}, {Gunn},
  {Harding}, {Johnston}, {Kauffmann}, {Kinemuchi}, {Klaene}, {Knapen},
  {Leauthaud}, {Li}, {Lin}, {Maiolino}, {Malanushenko}, {Malanushenko}, {Mao},
  {Maraston}, {McDermid}, {Merrifield}, {Nichol}, {Oravetz}, {Pan}, {Parejko},
  {Sanchez}, {Schlegel}, {Simmons}, {Steele}, {Steinmetz}, {Thanjavur},
  {Thompson}, {Tinker}, {van den Bosch}, {Westfall}, {Wilkinson}, {Wright},
  {Xiao}, \& {Zhang}}]{Bundy2015}
{Bundy}, K., {Bershady}, M.~A., {Law}, D.~R., {et~al.} 2015, \apj, 798, 7

\bibitem[{{Burns} {et~al.}(2014){Burns}, {Stritzinger}, {Phillips}, {Hsiao},
  {Contreras}, {Persson}, {Folatelli}, {Boldt}, {Campillay}, {Castell{\'o}n},
  {Freedman}, {Madore}, {Morrell}, {Salgado}, \& {Suntzeff}}]{Burns2014}
{Burns}, C.~R., {Stritzinger}, M., {Phillips}, M.~M., {et~al.} 2014, \apj, 789,
  32

\bibitem[{{Casasola} {et~al.}(2020){Casasola}, {Bianchi}, {De Vis}, {Magrini},
  {Corbelli}, {Clark}, {Fritz}, {Nersesian}, {Viaene}, {Baes}, {Cassar{\`a}},
  {Davies}, {De Looze}, {Dobbels}, {Galametz}, {Galliano}, {Jones}, {Madden},
  {Mosenkov}, {Tr{\v{c}}ka}, \& {Xilouris}}]{Casasola2020}
{Casasola}, V., {Bianchi}, S., {De Vis}, P., {et~al.} 2020, \aap, 633, A100

\bibitem[{{Casasola} {et~al.}(2017){Casasola}, {Cassar{\`a}}, {Bianchi},
  {Verstocken}, {Xilouris}, {Magrini}, {Smith}, {De Looze}, {Galametz},
  {Madden}, {Baes}, {Clark}, {Davies}, {De Vis}, {Evans}, {Fritz}, {Galliano},
  {Jones}, {Mosenkov}, {Viaene}, \& {Ysard}}]{casasola2017radial}
{Casasola}, V., {Cassar{\`a}}, L.~P., {Bianchi}, S., {et~al.} 2017, \aap, 605,
  A18

\bibitem[{{Cort{\^e}s} \& {Liddle}(2025)}]{Cortes2025}
{Cort{\^e}s}, M. \& {Liddle}, A.~R. 2025, \mnras, 544, L121

\bibitem[{{Cortese} {et~al.}(2012){Cortese}, {Ciesla}, {Boselli}, {Bianchi},
  {Gomez}, {Smith}, {Bendo}, {Eales}, {Pohlen}, {Baes}, {Corbelli}, {Davies},
  {Hughes}, {Hunt}, {Madden}, {Pierini}, {di Serego Alighieri}, {Zibetti},
  {Boquien}, {Clements}, {Cooray}, {Galametz}, {Magrini}, {Pappalardo},
  {Spinoglio}, \& {Vlahakis}}]{Cortese2012}
{Cortese}, L., {Ciesla}, L., {Boselli}, A., {et~al.} 2012, \aap, 540, A52

\bibitem[{{Davies} {et~al.}(2017){Davies}, {Baes}, {Bianchi}, {Jones},
  {Madden}, {Xilouris}, {Bocchio}, {Casasola}, {Cassara}, {Clark}, {De Looze},
  {Evans}, {Fritz}, {Galametz}, {Galliano}, {Lianou}, {Mosenkov}, {Smith},
  {Verstocken}, {Viaene}, {Vika}, {Wagle}, \& {Ysard}}]{Davies2017}
{Davies}, J.~I., {Baes}, M., {Bianchi}, S., {et~al.} 2017, \pasp, 129, 044102

\bibitem[{{de Vaucouleurs}(1974)}]{DeVa1974}
{de Vaucouleurs}, G. 1974, in IAU Symposium, Vol.~58, The Formation and
  Dynamics of Galaxies, ed. J.~R. {Shakeshaft}, 1

\bibitem[{{DES Collaboration} {et~al.}(2024){DES Collaboration}, {Abbott},
  {Acevedo}, {Aguena}, {Alarcon}, {Allam}, {Alves}, {Amon}, {Andrade-Oliveira},
  {Annis}, {Armstrong}, {Asorey}, {Avila}, {Bacon}, {Bassett}, {Bechtol},
  {Bernardinelli}, {Bernstein}, {Bertin}, {Blazek}, {Bocquet}, {Brooks},
  {Brout}, {Buckley-Geer}, {Burke}, {Camacho}, {Camilleri}, {Campos}, {Carnero
  Rosell}, {Carollo}, {Carr}, {Carretero}, {Castander}, {Cawthon}, {Chang},
  {Chen}, {Choi}, {Conselice}, {Costanzi}, {da Costa}, {Crocce}, {Davis},
  {DePoy}, {Desai}, {Diehl}, {Dixon}, {Dodelson}, {Doel}, {Doux},
  {Drlica-Wagner}, {Elvin-Poole}, {Everett}, {Ferrero}, {Fert{\'e}},
  {Flaugher}, {Foley}, {Fosalba}, {Friedel}, {Frieman}, {Frohmaier}, {Galbany},
  {Garc{\'\i}a-Bellido}, {Gatti}, {Gaztanaga}, {Giannini}, {Glazebrook},
  {Graur}, {Gruen}, {Gruendl}, {Gutierrez}, {Hartley}, {Herner}, {Hinton},
  {Hollowood}, {Honscheid}, {Huterer}, {Jain}, {James}, {Jeffrey}, {Kasai},
  {Kelsey}, {Kent}, {Kessler}, {Kim}, {Kirshner}, {Kovacs}, {Kuehn}, {Lahav},
  {Lee}, {Lee}, {Lewis}, {Li}, {Lidman}, {Lin}, {Malik}, {Marshall}, {Martini},
  {Mena-Fern{\'a}ndez}, {Menanteau}, {Miquel}, {Mohr}, {Mould}, {Muir},
  {M{\"o}ller}, {Neilsen}, {Nichol}, {Nugent}, {Ogando}, {Palmese}, {Pan},
  {Paterno}, {Percival}, {Pereira}, {Pieres}, {Malag{\'o}n}, {Popovic},
  {Porredon}, {Prat}, {Qu}, {Raveri}, {Rodr{\'\i}guez-Monroy}, {Romer},
  {Roodman}, {Rose}, {Sako}, {Sanchez}, {Sanchez Cid}, {Schubnell}, {Scolnic},
  {Sevilla-Noarbe}, {Shah}, {Smith}, {Smith}, {Soares-Santos}, {Suchyta},
  {Sullivan}, {Suntzeff}, {Swanson}, {S{\'a}nchez}, {Tarle}, {Taylor},
  {Thomas}, {To}, {Toy}, {Troxel}, {Tucker}, {Tucker}, {Uddin}, {Vincenzi},
  {Walker}, {Weaverdyck}, {Wechsler}, {Weller}, {Wester}, {Wiseman},
  {Yamamoto}, {Yuan}, {Zhang}, \& {Zhang}}]{DESSN2024}
{DES Collaboration}, {Abbott}, T.~M.~C., {Acevedo}, M., {et~al.} 2024, \apjl,
  973, L14

\bibitem[{{Dhawan} {et~al.}(2025){Dhawan}, {Popovic}, \& {Goobar}}]{Dhawan2025}
{Dhawan}, S., {Popovic}, B., \& {Goobar}, A. 2025, \mnras, 540, 1626

\bibitem[{{di Serego Alighieri} {et~al.}(2013){di Serego Alighieri}, {Bianchi},
  {Pappalardo}, {Zibetti}, {Auld}, {Baes}, {Bendo}, {Corbelli}, {Davies},
  {Davis}, {De Looze}, {Fritz}, {Gavazzi}, {Giovanardi}, {Grossi}, {Hunt},
  {Magrini}, {Pierini}, \& {Xilouris}}]{Alighieri2013}
{di Serego Alighieri}, S., {Bianchi}, S., {Pappalardo}, C., {et~al.} 2013,
  \aap, 552, A8

\bibitem[{{Draine} {et~al.}(2007){Draine}, {Dale}, {Bendo}, {Gordon}, {Smith},
  {Armus}, {Engelbracht}, {Helou}, {Kennicutt}, {Li}, {Roussel}, {Walter},
  {Calzetti}, {Moustakas}, {Murphy}, {Rieke}, {Bot}, {Hollenbach}, {Sheth}, \&
  {Teplitz}}]{Draine2007}
{Draine}, B.~T., {Dale}, D.~A., {Bendo}, G., {et~al.} 2007, \apj, 663, 866

\bibitem[{{Fitzpatrick} \& {Massa}(2007)}]{Fitzpatrick2007}
{Fitzpatrick}, E.~L. \& {Massa}, D. 2007, \apj, 663, 320

\bibitem[{{Foreman-Mackey} {et~al.}(2013){Foreman-Mackey}, {Hogg}, {Lang}, \&
  {Goodman}}]{emcee}
{Foreman-Mackey}, D., {Hogg}, D.~W., {Lang}, D., \& {Goodman}, J. 2013, \pasp,
  125, 306

\bibitem[{{Galliano} {et~al.}(2018){Galliano}, {Galametz}, \&
  {Jones}}]{Galliano2018}
{Galliano}, F., {Galametz}, M., \& {Jones}, A.~P. 2018, \araa, 56, 673

\bibitem[{{Ginolin} {et~al.}(2025){Ginolin}, {Rigault}, {Copin}, {Popovic},
  {Dimitriadis}, {Goobar}, {Johansson}, {Maguire}, {Nordin}, {Smith}, {Aubert},
  {Barjou-Delayre}, {Burgaz}, {Carreres}, {Dhawan}, {Deckers}, {Feinstein},
  {Fouchez}, {Galbany}, {Ganot}, {de Jaeger}, {Kim}, {Kuhn}, {Lacroix},
  {M{\"u}ller-Bravo}, {Nugent}, {Racine}, {Rosnet}, {Rosselli}, {Ruppin},
  {Sollerman}, {Terwel}, {Townsend}, {Dekany}, {Graham}, {Kasliwal}, {Groom},
  {Purdum}, {Rusholme}, \& {van der Walt}}]{Ginolin2025}
{Ginolin}, M., {Rigault}, M., {Copin}, Y., {et~al.} 2025, \aap, 694, A4

\bibitem[{{Gonz{\'a}lez-Gait{\'a}n} {et~al.}(2011){Gonz{\'a}lez-Gait{\'a}n},
  {Perrett}, {Sullivan}, {Conley}, {Howell}, {Carlberg}, {Astier}, {Balam},
  {Balland}, {Basa}, {Fouchez}, {Guy}, {Hardin}, {Hook.}, {Pain}, {Pritchet},
  {Regnault}, {Rich}, \& {Lidman}}]{Gonzalez2011}
{Gonz{\'a}lez-Gait{\'a}n}, S., {Perrett}, K., {Sullivan}, M., {et~al.} 2011,
  \apj, 727, 107

\bibitem[{{Goobar}(2008)}]{Goobar2008}
{Goobar}, A. 2008, \apjl, 686, L103

\bibitem[{{Goudfrooij} \& {de Jong}(1995)}]{Goudfrooij1995}
{Goudfrooij}, P. \& {de Jong}, T. 1995, \aap, 298, 784

\bibitem[{{Goudfrooij} {et~al.}(1994){Goudfrooij}, {de Jong}, {Hansen}, \&
  {Norgaard-Nielsen}}]{Goudfrooij1994}
{Goudfrooij}, P., {de Jong}, T., {Hansen}, L., \& {Norgaard-Nielsen}, H.~U.
  1994, \mnras, 271, 833

\bibitem[{{Graham} \& {Worley}(2008)}]{Graham2008}
{Graham}, A.~W. \& {Worley}, C.~C. 2008, \mnras, 388, 1708

\bibitem[{{Guy} {et~al.}(2005){Guy}, {Astier}, {Nobili}, {Regnault}, \&
  {Pain}}]{Guy2005}
{Guy}, J., {Astier}, P., {Nobili}, S., {Regnault}, N., \& {Pain}, R. 2005,
  \aap, 443, 781

\bibitem[{{Hakobyan} {et~al.}(2012){Hakobyan}, {Adibekyan}, {Aramyan},
  {Petrosian}, {Gomes}, {Mamon}, {Kunth}, \& {Turatto}}]{Hakobyan2012}
{Hakobyan}, A.~A., {Adibekyan}, V.~Z., {Aramyan}, L.~S., {et~al.} 2012, \aap,
  544, A81

\bibitem[{{Hakobyan} {et~al.}(2017){Hakobyan}, {Barkhudaryan}, {Karapetyan},
  {Mamon}, {Kunth}, {Adibekyan}, {Aramyan}, {Petrosian}, \&
  {Turatto}}]{hakobyan2017supernovae}
{Hakobyan}, A.~A., {Barkhudaryan}, L.~V., {Karapetyan}, A.~G., {et~al.} 2017,
  \mnras, 471, 1390

\bibitem[{{Hatano} {et~al.}(1998){Hatano}, {Branch}, \& {Deaton}}]{Hatano1998}
{Hatano}, K., {Branch}, D., \& {Deaton}, J. 1998, \apj, 502, 177

\bibitem[{{Holwerda} {et~al.}(2015{\natexlab{a}}){Holwerda}, {Keel},
  {Kenworthy}, \& {Mack}}]{Holwerda2015b}
{Holwerda}, B.~W., {Keel}, W.~C., {Kenworthy}, M.~A., \& {Mack}, K.~J.
  2015{\natexlab{a}}, \mnras, 451, 2390

\bibitem[{{Holwerda} {et~al.}(2015{\natexlab{b}}){Holwerda}, {Reynolds},
  {Smith}, \& {Kraan-Korteweg}}]{Holwerda2015a}
{Holwerda}, B.~W., {Reynolds}, A., {Smith}, M., \& {Kraan-Korteweg}, R.~C.
  2015{\natexlab{b}}, \mnras, 446, 3768

\bibitem[{{Hyde} \& {Bernardi}(2009)}]{hyde2009curvature}
{Hyde}, J.~B. \& {Bernardi}, M. 2009, \mnras, 394, 1978

\bibitem[{{Jha} {et~al.}(2007){Jha}, {Riess}, \& {Kirshner}}]{Jha2007}
{Jha}, S., {Riess}, A.~G., \& {Kirshner}, R.~P. 2007, \apj, 659, 122

\bibitem[{{Kessler} {et~al.}(2009){Kessler}, {Becker}, {Cinabro}, {Vanderplas},
  {Frieman}, {Marriner}, {Davis}, {Dilday}, {Holtzman}, {Jha}, {Lampeitl},
  {Sako}, {Smith}, {Zheng}, {Nichol}, {Bassett}, {Bender}, {Depoy}, {Doi},
  {Elson}, {Filippenko}, {Foley}, {Garnavich}, {Hopp}, {Ihara}, {Ketzeback},
  {Kollatschny}, {Konishi}, {Marshall}, {McMillan}, {Miknaitis}, {Morokuma},
  {M{\"o}rtsell}, {Pan}, {Prieto}, {Richmond}, {Riess}, {Romani}, {Schneider},
  {Sollerman}, {Takanashi}, {Tokita}, {van der Heyden}, {Wheeler}, {Yasuda}, \&
  {York}}]{Kessler2009}
{Kessler}, R., {Becker}, A.~C., {Cinabro}, D., {et~al.} 2009, \apjs, 185, 32

\bibitem[{{Larison} {et~al.}(2024){Larison}, {Jha}, {Kwok}, \&
  {Camacho-Neves}}]{Larison2024}
{Larison}, C., {Jha}, S.~W., {Kwok}, L.~A., \& {Camacho-Neves}, Y. 2024, \apj,
  961, 185

\bibitem[{{Laurikainen} {et~al.}(2007){Laurikainen}, {Salo}, {Buta}, \&
  {Knapen}}]{Laurikainen2007}
{Laurikainen}, E., {Salo}, H., {Buta}, R., \& {Knapen}, J.~H. 2007, \mnras,
  381, 401

\bibitem[{{Laurikainen} {et~al.}(2010){Laurikainen}, {Salo}, {Buta}, {Knapen},
  \& {Comer{\'o}n}}]{Laurikainen2010}
{Laurikainen}, E., {Salo}, H., {Buta}, R., {Knapen}, J.~H., \& {Comer{\'o}n},
  S. 2010, \mnras, 405, 1089

\bibitem[{{Legnardi} {et~al.}(2023){Legnardi}, {Milone}, {Cordoni}, {Lagioia},
  {Dondoglio}, {Marino}, {Jang}, {Mohandasan}, \& {Ziliotto}}]{Legnardi2023}
{Legnardi}, M.~V., {Milone}, A.~P., {Cordoni}, G., {et~al.} 2023, \mnras, 522,
  367

\bibitem[{{Le{\'s}niewska} {et~al.}(2023){Le{\'s}niewska}, {Micha{\l}owski},
  {Gall}, {Hjorth}, {Nadolny}, {Ryzhov}, \& {Solar}}]{Lesniewska2023}
{Le{\'s}niewska}, A., {Micha{\l}owski}, M.~J., {Gall}, C., {et~al.} 2023, \apj,
  953, 27

\bibitem[{{LSST Science Collaboration} {et~al.}(2009){LSST Science
  Collaboration}, {Abell}, {Allison}, {Anderson}, {Andrew}, {Angel}, {Armus},
  {Arnett}, {Asztalos}, {Axelrod}, \& et~al.}]{LSST2009}
{LSST Science Collaboration}, {Abell}, P.~A., {Allison}, J., {et~al.} 2009,
  ArXiv e-prints, 0912.0201 [\eprint[arXiv]{0912.0201}]

\bibitem[{{Makarov} {et~al.}(2014){Makarov}, {Prugniel}, {Terekhova},
  {Courtois}, \& {Vauglin}}]{Makarov2014}
{Makarov}, D., {Prugniel}, P., {Terekhova}, N., {Courtois}, H., \& {Vauglin},
  I. 2014, \aap, 570, A13

\bibitem[{{Mandel} {et~al.}(2011){Mandel}, {Narayan}, \&
  {Kirshner}}]{Mandel2011}
{Mandel}, K.~S., {Narayan}, G., \& {Kirshner}, R.~P. 2011, \apj, 731, 120

\bibitem[{{Mandel} {et~al.}(2017){Mandel}, {Scolnic}, {Shariff}, {Foley}, \&
  {Kirshner}}]{Mandel2017}
{Mandel}, K.~S., {Scolnic}, D.~M., {Shariff}, H., {Foley}, R.~J., \&
  {Kirshner}, R.~P. 2017, \apj, 842, 93

\bibitem[{{Mandel} {et~al.}(2022){Mandel}, {Thorp}, {Narayan}, {Friedman}, \&
  {Avelino}}]{Mandel2022}
{Mandel}, K.~S., {Thorp}, S., {Narayan}, G., {Friedman}, A.~S., \& {Avelino},
  A. 2022, \mnras, 510, 3939

\bibitem[{{Mannucci} {et~al.}(2005){Mannucci}, {Della Valle}, {Panagia},
  {Cappellaro}, {Cresci}, {Maiolino}, {Petrosian}, \& {Turatto}}]{Mannucci2005}
{Mannucci}, F., {Della Valle}, M., {Panagia}, N., {et~al.} 2005, \aap, 433, 807

\bibitem[{{Maoz} {et~al.}(2012){Maoz}, {Mannucci}, \& {Brandt}}]{Maoz2012}
{Maoz}, D., {Mannucci}, F., \& {Brandt}, T.~D. 2012, \mnras, 426, 3282

\bibitem[{{Micha{\l}owski} {et~al.}(2019){Micha{\l}owski}, {Hjorth}, {Gall},
  {Frayer}, {Tsai}, {Hirashita}, {Rowlands}, {Takeuchi}, {Le{\'s}niewska},
  {Behrendt}, {Bourne}, {Hughes}, {Spring}, {Zavala}, \&
  {Bartczak}}]{Michalowski2019}
{Micha{\l}owski}, M.~J., {Hjorth}, J., {Gall}, C., {et~al.} 2019, \aap, 632,
  A43

\bibitem[{{Mosenkov} {et~al.}(2022){Mosenkov}, {Usachev}, {Shakespear},
  {Guerrette}, {Baes}, {Bianchi}, {Xilouris}, {Gontcharov}, {Il'in}, {Marchuk},
  {Savchenko}, \& {Smirnov}}]{mosenkov2022distribution}
{Mosenkov}, A.~V., {Usachev}, P.~A., {Shakespear}, Z., {et~al.} 2022, \mnras,
  515, 5698

\bibitem[{{Newman} {et~al.}(2025){Newman}, {Larison}, {Jha}, {McQuinn},
  {Skillman}, {Dolphin}, {Dai}, {Howell}, {McCully}, {Bostroem}, {Hiramatsu},
  {Pellegrino}, \& {Padilla Gonzalez}}]{Newman2025}
{Newman}, M. J.~B., {Larison}, C., {Jha}, S.~W., {et~al.} 2025, arXiv e-prints,
  arXiv:2508.20023

\bibitem[{{Pastrav}(2021)}]{Pastrav2021}
{Pastrav}, B.~A. 2021, \mnras, 506, 452

\bibitem[{{Patil} {et~al.}(2007){Patil}, {Pandey}, {Sahu}, \&
  {Kembhavi}}]{Patil2007}
{Patil}, M.~K., {Pandey}, S.~K., {Sahu}, D.~K., \& {Kembhavi}, A. 2007, \aap,
  461, 103

\bibitem[{{Perlmutter} {et~al.}(1999){Perlmutter}, {Aldering}, {Goldhaber},
  {Knop}, {Nugent}, {Castro}, {Deustua}, {Fabbro}, {Goobar}, {Groom}, {Hook},
  {Kim}, {Kim}, {Lee}, {Nunes}, {Pain}, {Pennypacker}, {Quimby}, {Lidman},
  {Ellis}, {Irwin}, {McMahon}, {Ruiz-Lapuente}, {Walton}, {Schaefer}, {Boyle},
  {Filippenko}, {Matheson}, {Fruchter}, {Panagia}, {Newberg}, {Couch}, \&
  {Project}}]{Per1999}
{Perlmutter}, S., {Aldering}, G., {Goldhaber}, G., {et~al.} 1999, \apj, 517,
  565

\bibitem[{{Pilyugin} {et~al.}(2021){Pilyugin}, {Zinchenko}, {Lara-L{\'o}pez},
  {Nefedyev}, \& {V{\'\i}lchez}}]{Pilyugin2021}
{Pilyugin}, L.~S., {Zinchenko}, I.~A., {Lara-L{\'o}pez}, M.~A., {Nefedyev},
  Y.~A., \& {V{\'\i}lchez}, J.~M. 2021, \aap, 646, A54

\bibitem[{{Popovic} {et~al.}(2023){Popovic}, {Brout}, {Kessler}, \&
  {Scolnic}}]{Popovic2021}
{Popovic}, B., {Brout}, D., {Kessler}, R., \& {Scolnic}, D. 2023, \apj, 945, 84

\bibitem[{{Pritchet} {et~al.}(2024){Pritchet}, {Thanjavur}, {Bottrell}, \&
  {Gao}}]{Pritchet2024}
{Pritchet}, C., {Thanjavur}, K., {Bottrell}, C., \& {Gao}, Y. 2024, \aj, 167,
  131

\bibitem[{{Prugniel} \& {Simien}(1997)}]{Prugniel1997}
{Prugniel}, P. \& {Simien}, F. 1997, \aap, 321, 111

\bibitem[{{Pruzhinskaya} {et~al.}(2020){Pruzhinskaya}, {Novinskaya}, {Pauna},
  \& {Rosnet}}]{Pruzhinskaya2020}
{Pruzhinskaya}, M.~V., {Novinskaya}, A.~K., {Pauna}, N., \& {Rosnet}, P. 2020,
  \mnras, 499, 5121

\bibitem[{{Riello} \& {Patat}(2005)}]{riello2005extinction}
{Riello}, M. \& {Patat}, F. 2005, \mnras, 362, 671

\bibitem[{{Riess} {et~al.}(1998){Riess}, {Filippenko}, {Challis},
  {Clocchiatti}, {Diercks}, {Garnavich}, {Gilliland}, {Hogan}, {Jha},
  {Kirshner}, {Leibundgut}, {Phillips}, {Reiss}, {Schmidt}, {Schommer},
  {Smith}, {Spyromilio}, {Stubbs}, {Suntzeff}, \& {Tonry}}]{Rie1998}
{Riess}, A.~G., {Filippenko}, A.~V., {Challis}, P., {et~al.} 1998, \aj, 116,
  1009

\bibitem[{{Riess} {et~al.}(2022){Riess}, {Yuan}, {Macri}, {Scolnic}, {Brout},
  {Casertano}, {Jones}, {Murakami}, {Anand}, {Breuval}, {Brink}, {Filippenko},
  {Hoffmann}, {Jha}, {D'arcy Kenworthy}, {Mackenty}, {Stahl}, \&
  {Zheng}}]{Riess2022}
{Riess}, A.~G., {Yuan}, W., {Macri}, L.~M., {et~al.} 2022, \apjl, 934, L7

\bibitem[{{Rigault} {et~al.}(2025){Rigault}, {Smith}, {Goobar}, {Maguire},
  {Dimitriadis}, {Johansson}, {Nordin}, {Burgaz}, {Dhawan}, {Sollerman},
  {Regnault}, {Kowalski}, {Nugent}, {Andreoni}, {Amenouche}, {Aubert},
  {Barjou-Delayre}, {Bautista}, {Bellm}, {Betoule}, {Bloom}, {Carreres},
  {Chen}, {Copin}, {Deckers}, {de Jaeger}, {Feinstein}, {Fouchez}, {Fremling},
  {Galbany}, {Ginolin}, {Graham}, {Groom}, {Harvey}, {Kasliwal}, {Kenworthy},
  {Kim}, {Kuhn}, {Kulkarni}, {Lacroix}, {Laher}, {Masci}, {M{\"u}ller-Bravo},
  {Miller}, {Osman}, {Perley}, {Popovic}, {Purdum}, {Qin}, {Racine}, {Reusch},
  {Riddle}, {Rosnet}, {Rosselli}, {Ruppin}, {Senzel}, {Rusholme}, {Schweyer},
  {Terwel}, {Townsend}, {Tzanidakis}, {Wold}, \& {Yan}}]{Rigault2025}
{Rigault}, M., {Smith}, M., {Goobar}, A., {et~al.} 2025, \aap, 694, A1

\bibitem[{{Rino-Silvestre} {et~al.}(2025){Rino-Silvestre},
  {Gonz{\'a}lez-Gait{\'a}n}, {Mour{\~a}o}, {Duarte}, \&
  {Pereira}}]{Rino-Silvestre2025}
{Rino-Silvestre}, J., {Gonz{\'a}lez-Gait{\'a}n}, S., {Mour{\~a}o}, A.,
  {Duarte}, J., \& {Pereira}, B. 2025, arXiv e-prints, arXiv:2502.09875

\bibitem[{{Rose} {et~al.}(2022){Rose}, {Popovic}, {Scolnic}, \&
  {Brout}}]{Rose2022}
{Rose}, B.~M., {Popovic}, B., {Scolnic}, D., \& {Brout}, D. 2022, \mnras, 516,
  4822

\bibitem[{{Rowlands} {et~al.}(2012){Rowlands}, {Dunne}, {Maddox}, {Bourne},
  {Gomez}, {Kaviraj}, {Bamford}, {Brough}, {Charlot}, {da Cunha}, {Driver},
  {Eales}, {Hopkins}, {Kelvin}, {Nichol}, {Sansom}, {Sharp}, {Smith}, {Temi},
  {van der Werf}, {Baes}, {Cava}, {Cooray}, {Croom}, {Dariush}, {de Zotti},
  {Dye}, {Fritz}, {Hopwood}, {Ibar}, {Ivison}, {Liske}, {Loveday}, {Madore},
  {Norberg}, {Popescu}, {Rigby}, {Robotham}, {Rodighiero}, {Seibert}, \&
  {Tuffs}}]{Rowlands2012}
{Rowlands}, K., {Dunne}, L., {Maddox}, S., {et~al.} 2012, \mnras, 419, 2545

\bibitem[{{Rubin} {et~al.}(2025){Rubin}, {Aldering}, {Betoule}, {Fruchter},
  {Huang}, {Kim}, {Lidman}, {Linder}, {Perlmutter}, {Ruiz-Lapuente}, \&
  {Suzuki}}]{Rubin2023}
{Rubin}, D., {Aldering}, G., {Betoule}, M., {et~al.} 2025, \apj, 986, 231

\bibitem[{{Salim} {et~al.}(2018){Salim}, {Boquien}, \& {Lee}}]{Salim2018}
{Salim}, S., {Boquien}, M., \& {Lee}, J.~C. 2018, \apj, 859, 11

\bibitem[{{Scannapieco} \& {Bildsten}(2005)}]{Scannapieco2005}
{Scannapieco}, E. \& {Bildsten}, L. 2005, \apjl, 629, L85

\bibitem[{{Schlafly} {et~al.}(2016){Schlafly}, {Meisner}, {Stutz},
  {Kainulainen}, {Peek}, {Tchernyshyov}, {Rix}, {Finkbeiner}, {Covey}, {Green},
  {Bell}, {Burgett}, {Chambers}, {Draper}, {Flewelling}, {Hodapp}, {Kaiser},
  {Magnier}, {Martin}, {Metcalfe}, {Wainscoat}, \& {Waters}}]{Schlafly2016}
{Schlafly}, E.~F., {Meisner}, A.~M., {Stutz}, A.~M., {et~al.} 2016, \apj, 821,
  78

\bibitem[{{Scolnic} {et~al.}(2015){Scolnic}, {Casertano}, {Riess}, {Rest},
  {Schlafly}, {Foley}, {Finkbeiner}, {Tang}, {Burgett}, {Chambers}, {Draper},
  {Flewelling}, {Hodapp}, {Huber}, {Kaiser}, {Kudritzki}, {Magnier},
  {Metcalfe}, \& {Stubbs}}]{Scolnic2015}
{Scolnic}, D., {Casertano}, S., {Riess}, A., {et~al.} 2015, \apj, 815, 117

\bibitem[{{S{\'e}rsic}(1963)}]{Sersic1963}
{S{\'e}rsic}, J.~L. 1963, Boletin de la Asociacion Argentina de Astronomia La
  Plata Argentina, 6, 41

\bibitem[{{Sersic}(1968)}]{Sersic1968}
{Sersic}, J.~L. 1968, {Atlas de Galaxias Australes}

\bibitem[{{Smith} {et~al.}(2016){Smith}, {Eales}, {De Looze}, {Baes}, {Bendo},
  {Bianchi}, {Boquien}, {Boselli}, {Buat}, {Ciesla}, {Clemens}, {Clements},
  {Cooray}, {Cortese}, {Davies}, {Fritz}, {Gomez}, {Hughes}, {Karczewski},
  {Lu}, {Oliver}, {Remy-Ruyer}, {Spinoglio}, \& {Viaene}}]{smith2016far}
{Smith}, M. W.~L., {Eales}, S.~A., {De Looze}, I., {et~al.} 2016, \mnras, 462,
  331

\bibitem[{{Smith} {et~al.}(2012){Smith}, {Gomez}, {Eales}, {Ciesla}, {Boselli},
  {Cortese}, {Bendo}, {Baes}, {Bianchi}, {Clemens}, {Clements}, {Cooray},
  {Davies}, {De Looze}, {di Serego Alighieri}, {Fritz}, {Gavazzi}, {Gear},
  {Madden}, {Mentuch}, {Panuzzo}, {Pohlen}, {Spinoglio}, {Verstappen},
  {Vlahakis}, {Wilson}, \& {Xilouris}}]{smith2012herschel}
{Smith}, M.~W.~L., {Gomez}, H.~L., {Eales}, S.~A., {et~al.} 2012, \apj, 748,
  123

\bibitem[{{Thorp} \& {Mandel}(2022)}]{Thorp2022}
{Thorp}, S. \& {Mandel}, K.~S. 2022, \mnras, 517, 2360

\bibitem[{{Thorp} {et~al.}(2024){Thorp}, {Mandel}, {Jones}, {Kirshner}, \&
  {Challis}}]{Thorp2024}
{Thorp}, S., {Mandel}, K.~S., {Jones}, D.~O., {Kirshner}, R.~P., \& {Challis},
  P.~M. 2024, \mnras, 530, 4016

\bibitem[{{Thorp} {et~al.}(2021){Thorp}, {Mandel}, {Jones}, {Ward}, \&
  {Narayan}}]{Thorp2021}
{Thorp}, S., {Mandel}, K.~S., {Jones}, D.~O., {Ward}, S.~M., \& {Narayan}, G.
  2021, \mnras, 508, 4310

\bibitem[{{Toy} {et~al.}(2025){Toy}, {Wiseman}, {Sullivan}, {Scolnic},
  {Vincenzi}, {Brout}, {Davis}, {Frohmaier}, {Galbany}, {Lidman}, {Lee},
  {Kelsey}, {Kessler}, {M{\"o}ller}, {Popovic}, {S{\'a}nchez}, {Shah}, {Smith},
  {Aguena}, {Allam}, {Alves}, {Bacon}, {Brooks}, {Burke}, {Rosell},
  {Carretero}, {da Costa}, {Pereira}, {Desai}, {Diehl}, {Doel},
  {Drlica-Wagner}, {Everett}, {Ferrero}, {Flaugher}, {Frieman},
  {Garc{\'\i}a-Bellido}, {Gatti}, {Gaztanaga}, {Giannini}, {Gruendl},
  {Gutierrez}, {Hinton}, {Hollowood}, {Honscheid}, {James}, {Kuehn}, {Lahav},
  {Lee}, {Marshall}, {Mena-Fern{\'a}ndez}, {Miquel}, {Palmese}, {Pieres},
  {Malag{\'o}n}, {Romer}, {Samuroff}, {Sanchez}, {Cid}, {Schubnell}, {Suchyta},
  {Swanson}, {Tarle}, {Tucker}, {Vikram}, {Walker}, \& {Weaverdyck}}]{Toy2025}
{Toy}, M., {Wiseman}, P., {Sullivan}, M., {et~al.} 2025, \mnras, 538, 181

\bibitem[{{Uddin} {et~al.}(2024){Uddin}, {Burns}, {Phillips}, {Suntzeff},
  {Freedman}, {Brown}, {Morrell}, {Hamuy}, {Krisciunas}, {Wang}, {Hsiao},
  {Goobar}, {Perlmutter}, {Lu}, {Stritzinger}, {Anderson}, {Ashall},
  {Hoeflich}, {Shappee}, {Persson}, {Piro}, {Baron}, {Contreras}, {Galbany},
  {Kumar}, {Shahbandeh}, {Davis}, {Anais}, {Busta}, {Campillay},
  {Castell{\'o}n}, {Corco}, {Diamond}, {Gall}, {Gonzalez}, {Holmbo}, {Roth},
  {Ser{\'o}n}, {Taddia}, {Torres}, {Baltay}, {Folatelli}, {Hadjiyska},
  {Kasliwal}, {Nugent}, {Rabinowitz}, \& {Ryder}}]{Uddin2024}
{Uddin}, S.~A., {Burns}, C.~R., {Phillips}, M.~M., {et~al.} 2024, \apj, 970, 72

\bibitem[{{Ward} {et~al.}(2023){Ward}, {Dhawan}, {Mandel}, {Grayling}, \&
  {Thorp}}]{Ward2023}
{Ward}, S.~M., {Dhawan}, S., {Mandel}, K.~S., {Grayling}, M., \& {Thorp}, S.
  2023, \mnras, 526, 5715

\bibitem[{{Weingartner} \& {Draine}(2001)}]{Weingartner2001ApJ}
{Weingartner}, J.~C. \& {Draine}, B.~T. 2001, \apj, 548, 296

\bibitem[{{Weinzirl} {et~al.}(2009){Weinzirl}, {Jogee}, {Khochfar}, {Burkert},
  \& {Kormendy}}]{Weinzirl2009}
{Weinzirl}, T., {Jogee}, S., {Khochfar}, S., {Burkert}, A., \& {Kormendy}, J.
  2009, \apj, 696, 411

\bibitem[{{Wojtak} \& {Hjorth}(2022)}]{Wojtak2022}
{Wojtak}, R. \& {Hjorth}, J. 2022, \mnras, 515, 2790

\bibitem[{{Wojtak} \& {Hjorth}(2024)}]{Wojtak2024}
{Wojtak}, R. \& {Hjorth}, J. 2024, \mnras, 533, 2319

\bibitem[{{Wojtak} \& {Hjorth}(2025)}]{Wojtak2025}
{Wojtak}, R. \& {Hjorth}, J. 2025, \aap, 702, A176

\bibitem[{{Wojtak} {et~al.}(2023){Wojtak}, {Hjorth}, \&
  {Hjortlund}}]{Wojtak2023}
{Wojtak}, R., {Hjorth}, J., \& {Hjortlund}, J.~O. 2023, \mnras, 525, 5187

\bibitem[{{Xilouris} {et~al.}(1999){Xilouris}, {Byun}, {Kylafis}, {Paleologou},
  \& {Papamastorakis}}]{Xilouris1999}
{Xilouris}, E.~M., {Byun}, Y.~I., {Kylafis}, N.~D., {Paleologou}, E.~V., \&
  {Papamastorakis}, J. 1999, \aap, 344, 868

\end{thebibliography}

\end{document}